\documentclass[a4paper, 11pt]{article}
\usepackage{graphicx}
\usepackage[usenames]{color}
\usepackage[flushleft]{threeparttable}
\DeclareGraphicsExtensions{.pdf,.png,.jpg,.mps,.eps,.ps}
\usepackage{amsmath,amssymb,gensymb}
\usepackage{natbib} 
\setcitestyle{authoryear,open={(},close={)}}
\usepackage{color}
\usepackage{lscape}
\usepackage{psfrag}
\usepackage[colorlinks,citecolor=blue,urlcolor=blue,bookmarks=false,hypertexnames=true,pagebackref=true]{hyperref}

\newcommand\nicer{{\it NICER}}
\newcommand\nustar{{\it NuSTAR}}

\newcommand\sax{{\it BeppoSAX}}

\newcommand\rxte{{\it RXTE}}
\newcommand\xmm{{\it XMM-Newton}}
\newcommand\inte{{\it INTEGRAL}}

\newcommand\kev{{\rm~keV}}

\newcommand\kms{\ifmmode {\rm~km\ s}^{-1} \else ~km s$^{-1}$\fi}
\newcommand\Hunit{\ifmmode {\rm~km\ s}^{-1}\ {\rm Mpc}^{-1}
        \else ~km s$^{-1}$ Mpc$^{-1}$\fi}
\newcommand\ctssec{\ifmmode {\rm~count\ s}^{-1} \else ~count s$^{-1}$\fi}
\newcommand\ergsec{\ifmmode {\rm~erg\ s}^{-1} \else
        ~erg s$^{-1}$\fi}
\newcommand\funit{\ifmmode {\rm~erg\ s}^{-1}\;{\rm cm}^{-2} \else
        ~ergs s$^{-1}$ cm$^{-2}$\fi}
\newcommand\phflux{\ifmmode {\rm~photon\ s}^{-1}\;{\rm cm}^{-2}
        \else   ~photon s$^{-1}$ cm$^{-2}$\fi}
\newcommand\efluxA{\ifmmode {\rm~erg\ s}^{-1}\;{\rm cm}^{-2}\;{\rm
        \AA}^{-1} \else ~erg s$^{-1}$ cm$^{-2}$ \AA$^{-1}$\fi}
\newcommand\efluxHz{\ifmmode {\rm~erg\ s}^{-1}\;{\rm cm}^{-2}\;{\rm
        Hz}^{-1} \else ~erg s$^{-1}$ cm$^{-2}$ Hz$^{-1}$\fi}
\newcommand\cc{\ifmmode {\rm~cm}^{-3} \else cm$^{-3}$\fi}
\newcommand\FWHM{\ifmmode {\rm~FWHM} \else ${\rm~FWHM}$\fi}
\newcommand\Msun{\ifmmode M_{\odot} \else $M_{\odot}$\fi}
\newcommand\Lsun{\ifmmode L_{\odot} \else $L_{\odot}$\fi}

\newcommand\hbeta{\ifmmode {\rm H}\beta \else H$\beta$\fi}
\newcommand\Kalpha{\ifmmode {\rm K}\alpha \else K$\alpha$\fi}
\newcommand\nh{\ifmmode N_{\rm H} \else N$_{\rm H}$\fi}

\title{\nustar{} discovers a long type-I X-ray burst from the clocked burster GS~1826-24}
\author{Aditya S. Mondal$^{1}\thanks{E-mail: adityas.mondal@visva-bharati.ac.in}$, Mayukh Pahari$^{2}$, Gulab C. Dewangan$^{3}$  \\
{\small
$^{1}$Department of physics, Visva-Bharati, Santiniketan, West Bengal, 731235, India} \\
{\small $^{2}$Department of Physics, Indian Institute of Technology Hyderabad, Hyderabad, Kandi, 502285 Sangareddy, India} \\
{\small $^{3}$Inter-University Centre for  Astronomy \& Astrophysics (IUCAA), Pune, 411007, India}\\
}

\date{\today}
\begin{document}
\maketitle
\begin{abstract}
The source GS~1826-24 is a neutron star low mass X-ray binary known as the “clocked burster” because of its extremely regular bursting behavior. We report on the detection of a long type-I X-ray burst from this source. We perform a detailed spectroscopic analysis of the long X-ray burst, lasting for $\sim 600$ s, seen in the \nustar{} observation carried out on  2022 September. The persistent emission is well described by an absorbed thermal Comptonization model {\tt nthcomp}, and the source exhibits a soft spectral state during this observation. The observed burst exhibits a rise time of $\sim 25$ s and a decay time of $\sim 282$ s. The time-resolved spectroscopy of the burst shows a significant departure from a pure thermal spectrum and is described with a model consisting of a varying-temperature blackbody plus an evolving persistent emission component. We observe a significant enhancement in the persistent emission during the burst. The enhancement of the pre-burst persistent flux is possibly due to Poynting-Robertson drag or coronal reprocessing. At the peak of the burst, the blackbody temperature and the blackbody emitting radius reached a maximum of $2.10\pm 0.07$ keV and $5.5\pm 2.1$ km, respectively. The peak flux ($F_{peak}$) during the burst is $\approx 2.4\times 10^{-8}$ ergs cm$^{-2}$ s$^{-1}$, which corresponds to a luminosity of $\approx 9.7\times 10^{37}$ ergs s$^{-1}$.

\end{abstract}

\section{Introduction}
Type I X-ray bursts are thermonuclear explosions on the surface layers of weakly magnetized neutron stars (NSs) accreting mass from a low-mass companion, which is usually a low-mass star ($<3M_{\odot}$) or a white dwarf (\citealt{1993SSRv...62..223L, 2006csxs.book..113S}). The NS accretes mainly hydrogen and/or helium from its Roche lobe filling companion star. The energy source of X-ray bursts is nuclear burning on the surface of NSs, where accreted hydrogen and helium from a companion star are converted into heavy nuclei \citep{2021ASSL..461..209G}. These bursts last from tens to hundreds of seconds, depending on the ignition depth and the composition of the burning material \citep{2000AIPC..522..359B}. Observed burst profiles show a rapid rise followed by an exponential decay, representing the cooling of the NS surface. Short type- I X-ray bursts, lasting up to tens of seconds, are attributed to the triple-$\alpha$ process in a helium-rich fuel. The longer bursts of up to a few hundred seconds are expected from hydrogen-rich or mixed H/He fuels (\citealt{1981ApJ...247..267F, 2000AIPC..522..359B, 2003ApJ...595.1077C}). During some highly energetic bursts, the radiation pressure may reach the Eddington limit, so the burning layer expands, lifting off from the surface, thus leading to the expansion of the photosphere of the NS (\citealt{2003ApJ...595.1077C, 2003A&A...399..663K}). During the thermonuclear burning, the supply of the seed photons from the surface of the NS is expected to increase and cool the Comptonized corona around the NS (\citealt{2003A&A...399.1151M, 2013ApJ...777L...9C, 2013MNRAS.432.2773J, 2020A&A...634A..58S}). The burst X-ray spectra are usually fit with a blackbody (kT peaking at 2–3 keV; \citealt{1977ApJ...212L..73S}) that cools down during the burst decay. To model the continuum burst emission, one usually assumes that the persistent emission remains unchanged during the burst, and one subtracts it off as part of the background (see, e.g., \citealt{2003A&A...399..663K}). However, \citet{2013ApJ...772...94W} proposed that the persistent emission intensity is variable, thus involving a multiplicative factor $f_{a}$ to the persistent emission (assuming an unchanged spectral shape) for fitting the spectrum during the burst. In this varying persistent flux model, the value of $f_{a}$ seems to be proportional to the intensity of bursts. \\

The source GS~1826-24 is a typical NS Low Mass X-ray Binary (LMXB), located at a distance of 5.7 kpc \citep{2016ApJ...818..135C}. The source was initially classified as a blackhole candidate because of its similarity with Cyg X-1. Later, the detection of the three type-I X-ray bursts with \sax{} established that the compact object in this system is a weakly magnetized NS \citep{1997IAUC.6611....1U}. This source GS~1826-24 was discovered as a new transient by Ginga in 1988 September \citep{1989ESASP.296....3T}. It has been X-ray bright and exhibiting regular type-I X-ray bursts since its discovery. The source GS~1826-24 is commonly known as the “clocked burster” because of its extremely regular bursting behavior with a burst recurrence time close to 6 hours (\citealt{1999ApJ...514L..27U, 2004ApJ...601..466G}). GS~1826-24 is a unique X-ray burster that shows remarkable agreement with theoretical models of recurrence times, energetics, and lightcurves. \citet{2004ApJ...601..466G} studied 24 bursts observed from GS~1826-24 by the \rxte{} between 1997 November and 2002 July and measured a relationship between persistent X-ray flux and burst recurrence time. The source GS~1826-24 remained in the persistent hard spectral state from 1996 until 2014, and the spectral state was characterized by a dominant power-law component \citep{2008ApJS..179..360G, 2011A&A...529A.155C, 2015PASJ...67...92A, 2016ApJ...817..101R}. The source exhibited a clocked burster state with burst recurrence times of 3 to 6 hours between 1997 and 2007, as reported by \citet{2008ApJ...681..506T}. Around 2013, the source gradually brightened and transitioned to the soft state after 2014 (Fig. 12 in \citealt{2022PASJ...74..974A}). \\

The source was not active in its clocked burster state, and the pattern of regular bursting behaviour has been broken, with the transition to weaker, irregular bursts accompanied by a soft spectral state, in 2014 June \citep{2016ApJ...818..135C}. Shortly after this soft episode, the source returned to its hard state again. According to observations by Swift/BAT and MAXI, the intensity of GS 1826-238 gradually decreased and transitioned back to the hard state \citep{2016ApJ...818..135C, 2018MNRAS.474.1583J}. After 2016, this source transfers into intermediate and soft states \citep{2018MNRAS.474.1583J}. There have been 15 \nustar{} StrayCat observations of GS~1826-24 between 2014 February and 2021 November, and a transition of the source from the island atoll state (hard) to a banana branch (soft) has been observed during this time interval \citep{2023ApJ...947...81Y}. Again, approximately after one year, \nustar{} observed the source in 2022 September in an extended soft spectral state, and here we present the analysis of this \nustar{} observation. Presumably, the change in the observed behavior is correlated to a change in the mass accretion rate onto the NS \citep{1989A&A...225...79H}. Recent observation suggests that GS~1826-238 may have entered again a clocked burster state for the first time in 10 years with a shortest burst recurrence period of approximately 1.6 hours \citep{2025ATel17245....1I}.\\

\citet{2016ApJ...818..135C} detected the Eddington limited type-I X-ray burst and derived the distance of the source to be $5.7\pm 0.2$ kpc, assuming an isotropic emission. An excess in the soft flux and a shortage in the hard X-ray flux was reported during the type-I X-ray burst observed in this source (\citealt{2014ApJ...782...40J, 2020A&A...634A..58S}). The source is also known for the long burst tails, lasting $\sim 100$ s, which were powered by rp-process hydrogen burning \citep{2004ApJ...601..466G}. The inferred accretion rate varies little on short timescales, and is within the range $5–13\% \dot{m}_{Edd}$ (\citealt{2008ApJS..179..360G, 2016ApJ...818..135C}). The \nicer{} observations of this source revealed 7–9 mHz oscillations with fractional rms 2 per cent at 6 keV \citep{2018AAS...23133303S}. The first superburst (lasted for at least 3 hours) from GS 1826–24 was detected by the MAXI/GSC on 2018 February \citep{2018ATel11422....1I}.\\

In the present work, we have carried out a detailed time-resolved spectral analysis of the long type-I X-ray burst observed during the \nustar{} observation. We have also performed the spectral analysis of the persistent emission. An enhancement of the persistent emission has been observed during the burst. We discuss this in detail. This paper is structured as follows. In Section 2, we describe the observations and data reduction used in this study, and in Section 3, we discuss the light curve of the burst. Section 4 describes the spectral analysis techniques adopted for the persistent and burst emission. Finally, we discuss the observational findings and their implications in Section 5.\\

\begin{figure*}
\centering
\includegraphics[scale=0.35, angle=0]{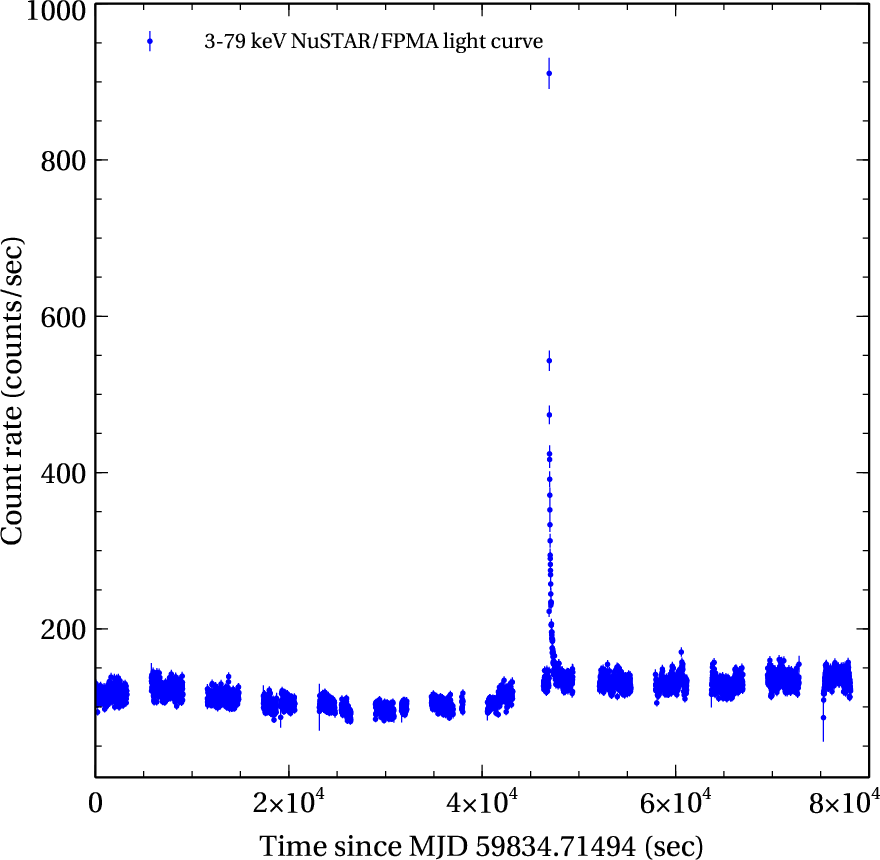}
\includegraphics[scale=0.35, angle=0]{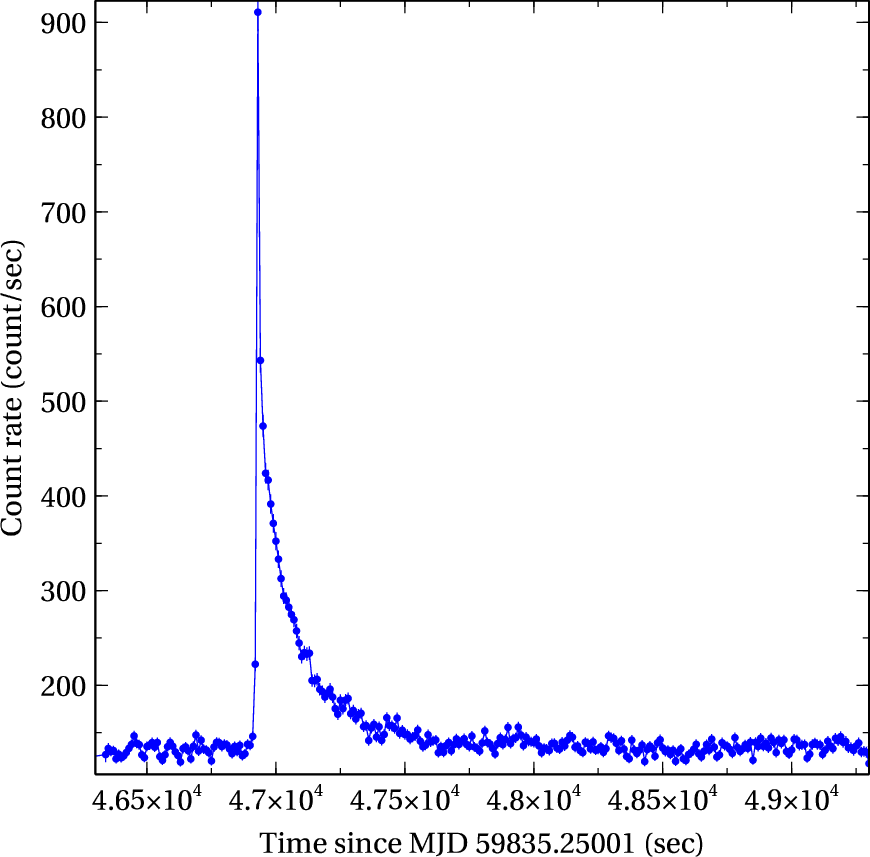}
\includegraphics[scale=0.35, angle=0]{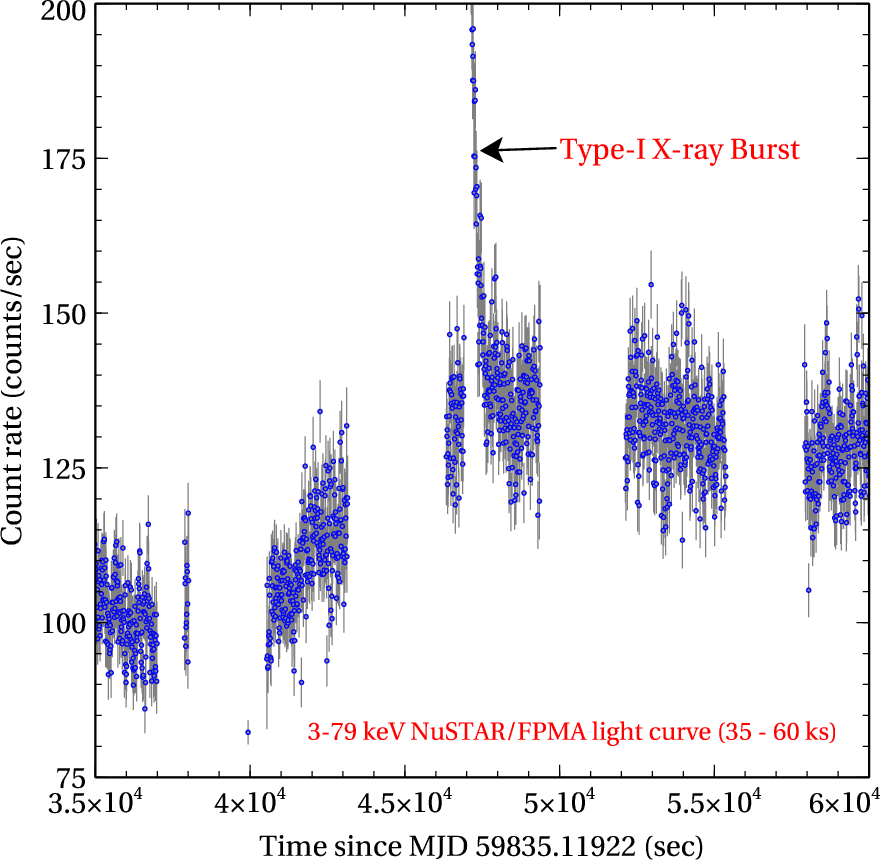}
\caption{Left panel: It shows the $3-79\kev{}$ \nustar{}/FPMA whole light curve of the source GS~1826-24. The source shows the presence of one thermonuclear X-ray burst at about $\sim 47$th ks of this observation. Middle panel: zoomed version of the light curve, only one orbit of \nustar{} observation is shown, where the burst has been observed. Right panel: It shows the persistent light curve within 35th to 60th ks of this observation. A change in the persistent flux has been observed from the \nustar{} orbit, where the burst is located.} 
\label{Fig1}
\end{figure*}

\section{observation and data reduction}
The Nuclear Spectroscopic Telescope Array (\nustar{}) is a hard X-ray (3–79 keV) focusing telescope with two identical, co-aligned telescopes, FPMA and FPMB. Its effective area peaks at $\approx 900$ cm$^{2}$ (adding up the two modules) around 10 keV, and its energy resolution at 10 keV is 400 eV \citep{2013ApJ...770..103H}. \nustar{} observed the source on 12th September 2022 (obs ID: 30801022002) for a total of $\sim 31$ ks. The current analysis is based only on this observation.  \\

We processed the \nustar{} data using the data analysis Software {\tt NuSTARDAS v2.1.5}, which is distributed with {\tt HEASOFT v6.35.1}, and used the latest calibration files {\tt v20250415} available during the analysis. We used the task {\tt nupipeline v0.4.12} to generate the calibrated and screened event ﬁles. A circular extraction region of $120^{''}$ radius was used to study the persistent and burst spectrum. To collect background events, we selected a region of similar size from the same chip but in a position away from the source. Spectra and light curves were extracted using the FTOOL {\tt nuproducts} from the FPMA and FPMB. During this observation, we created good time intervals (GTIs) for the persistent emission and the observed type-I X-ray burst. For the time-resolved spectroscopy, we created seven segments along the burst profile and built the corresponding GTI files. We then used them to extract the burst spectra for each segment. We found that the background dominates the signal at energies over $\geq 30$ keV. \\

\section{Light Curve}
\begin{figure}
\centering
\includegraphics[scale=0.60, angle=0]{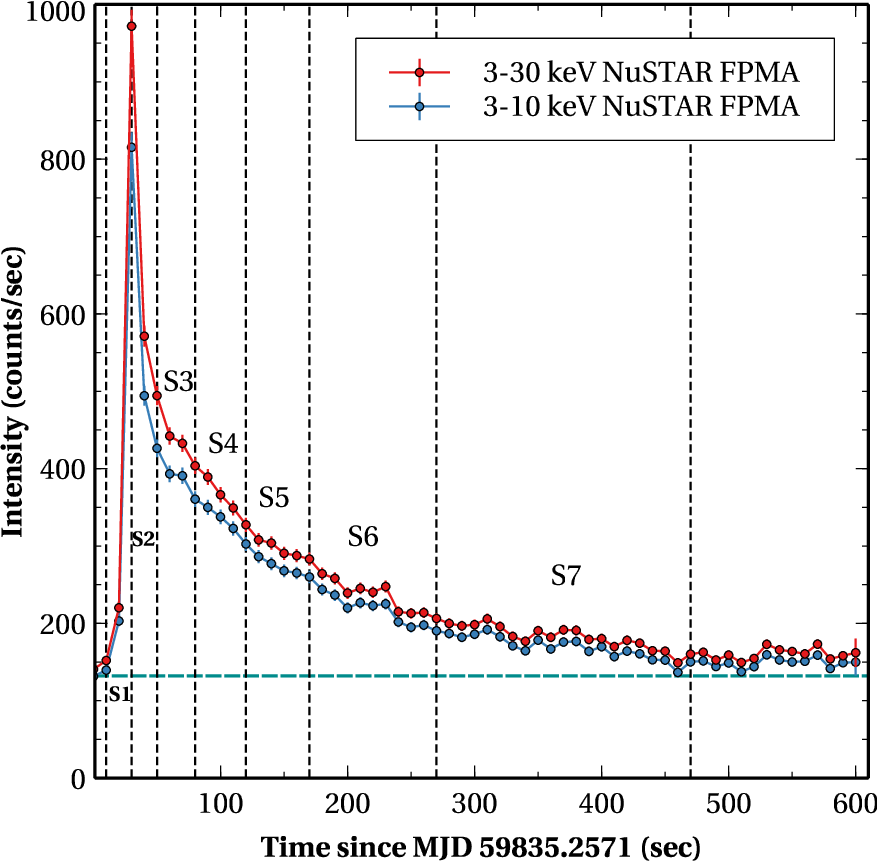}
\caption{Burst profiles in the different energy bands have been shown here. The burst interval is divided into seven segments ($\rm S1...\rm S7$). The integration time for each segment is 20s, 20s, 30s, 40s, 50s, 100s, 200s, respectively. The horizontal line (green) shows the persistent emission level.} 
\label{Fig2}
\end{figure}

In the left panel of Figure~\ref{Fig1}, we show the \nustar{}/FPMA light curve of the source in the $3-79$ \kev{} energy band. The average count rate during the \nustar{} observation was $\sim 100-130$ counts s$^{-1}$. One type-I thermonuclear X-ray burst is present at about $\sim 47$th ks (MJD  $59835.2571$) during the \nustar{} observation (see left and middle panel of Figure~\ref{Fig1}). From this particular \nustar{} orbit, a change in the persistent flux has been observed (see right panel of Figure~\ref{Fig1}). The X-ray burst has a typical sharp rise and near-exponential decay. The observed burst has a long tail, lasting $\sim 600$ s, and the peak of the burst seems to reach approximately $\sim 1000$ counts s$^{-1}$, about a factor of $\sim 10$ above the level of the persistent emission. Considering the rising and exponential decay time, the type-I burst lasts about $\sim 600$ s. We extracted the light curve (see Figure~\ref{Fig2}) of the burst in different energy bands. We performed the time-resolved spectral analysis of the burst to investigate the spectral properties in detail. For the persistent emission spectrum, we took a time interval of around $400$ s just before the rise of the burst and a time interval of $2$ ks before the burst, but from a different \nustar{} orbit.

\section{spectral analysis}
The persistent and the burst spectra are analyzed with the X-ray spectral package {\tt XSPEC} $v12.15.0$ \citep{1996ASPC..101...17A}. For the persistent emission, we model the \nustar{} FPMA and FPMB spectra simultaneously between $3$ to $30$\kev{} energy band. To account for cross-calibration of the two instruments, FPMA and FPMB, a constant multiplication factor {\tt constant} was included in the modeling. The value of {\tt constant} for FPMA was fixed to $1$, allowing it to vary for the FPMB. We used the {\tt TBabs} model to account for interstellar absorption along the line of sight with the {\tt wilm} abundances \citep{2000ApJ...542..914W} and the {\tt vern} \citep{1996ApJ...465..487V} photoelectric cross-section. Spectral uncertainties are quoted at the $1 \sigma$ confidence level. 

\subsection{Persistent spectral analysis}
We investigate the persistent emission before the burst to quantify the spectral shape and to measure the persistent flux for comparison with the burst spectrum. We generate a pre-burst spectrum of the source from two different time intervals: (1) a time interval of 2 ks (from the elapsed time 41st ks to 43rd ks) has been chosen before the burst but from a different \nustar{} orbit (2) another time interval of $400$ s has been chosen just before the burst onset depending on the data available in the same \nustar{} orbit. After generating the $3 - 30$ keV \nustar{} FPMA and FPMB persistent spectrum, we tried to fit them simultaneously with an absorbed cutoff power law model {\tt TBabs*cutoffpl}. Due to the lack of coverage below 3 keV, we fixed the hydrogen column density at $0.2\times 10^{22}$ cm$^{-2}$ \citep{1990ARA&A..28..215D}. We found that both persistent emission spectra are well described by the model {\tt tbabs*cutoffpl} with $\chi^2/dof=497/485$ and $\chi^2/dof=352/316$ for the two persistent spectra, respectively. The persistent emission is characterized by a power law component with a photon index of $\Gamma=0.93-1.20$ and the cutoff energy $E_{cut}\sim 4\kev{}$. This power law component can be explained as the result of inverse Compton scattering in the corona. However, the direct disc emission is not spectrally detected due to the low energy limitation of \nustar{}. \\

\begin{figure}
\centering
\includegraphics[scale=0.40, angle=-90]{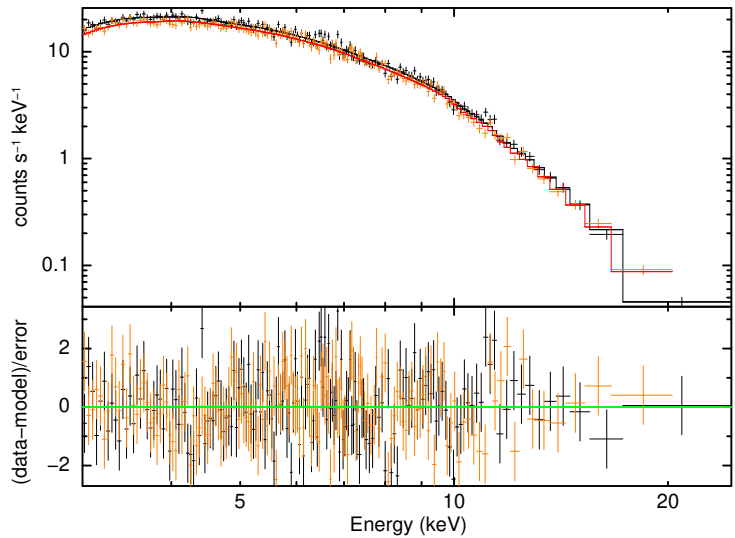}
\caption{Persistent \nustar{} FPMA (black) and FPMB (orange) spectrum of the source GS~1826-24, extracted from an exposure time of 400s just before the onset of the burst. The unfolded spectral data for the best-fit model {\tt const*TBabs*nthcomp} is shown. The lower panel of this plot shows the ratio of the data to the model in units of $\sigma$.  }
\label{Fig3}
\end{figure}

To describe the persistent emission more precisely, we employed a thermal Comptonization model {\tt nthcomp} \citep{1996MNRAS.283..193Z}, which attempts to simulate the upscattering of photons through the corona from a seed spectrum parameterized either by the inner temperature of the accretion disc or the NS surface/boundary layer. We then tested the physically-motivated Comptonization model {\tt nthcomp}, setting the photon seed input to a disc blackbody. This model {\tt TBabs*nthcomp} also describes the continuum emission very well with $\chi^2/dof=499/484$ and $\chi^2/dof=350/315$ for two persistent spectra, respectively (see Figure~\ref{Fig3}). We estimated the electron temperature at $kT_{e}= 2.40-2.60 \kev{}$ with photon index $\Gamma= 2.15-2.30$. The seed photon temperature is found to be low $kT_{seed}\sim 0.8-0.9\kev{}$, indicating the difficulties in detecting the soft disc blackbody component arising from the accretion disc. The electron temperature of the corona, $kT_{e}$, is roughly comparable to the observed cutoff energy ($\sim 4$\kev{}) of the spectrum. The best-fit parameters for the persistent spectrum are quoted in Table~\ref{table1}. \\

\begin{table*}
   \centering
\caption{Fit results (persistent emission): Best-fitting spectral parameters of the \nustar{} observation of the source GS~1826-24. For the persistent emission the models are Model1: {\tt const*TBabs*cutoffpl} and Model2: {\tt const*TBabs*nthcomp}} 
\begin{tabular}{|p{2.0cm}|p{3.0cm}|p{2.5cm}|p{2.5cm}|}
    \hline
    Component     & Parameter (unit) & Persistent (2ks) & Persistent (400s)\\
    \hline
    {\scshape Constant} & FPMB (wrt FPMA) & $0.99\pm 0.004$    &$0.98\pm 0.003$  \\
    {\scshape tbabs}    & $N_{H}$($\times 10^{22}\;\text{cm}^{-2}$) &  $0.20 $(f)  &$0.20 $(f)   \\
    {\scshape cutoffpl} & $\Gamma$ &  $1.19\pm 0.02$     &$1.01\pm 0.05$   \\
    & $\rm E_{cut}$ (keV) &  $3.97\pm 0.05$    &$3.70\pm 0.10$   \\
    & Norm &  $1.66\pm 0.03$    &$1.50\pm 0.07$    \\ 
    \hline
    & $\chi^{2}/dof$ &  $497/485$    &$352/316$  \\ 
    \hline
    {\scshape nthcomp} & $\Gamma$ &  $2.26\pm 0.03$     &$2.16\pm 0.07$   \\
    & $kT_{e}$ (keV) &    $2.52\pm 0.04$      &$2.40\pm 0.08$  \\
    & $kT_{bb}$ (keV) &  $0.84\pm 0.04$  &$0.86\pm 0.10$  \\
    & Norm & $0.94\pm 0.03$ &$0.98\pm 0.08$ \\
    {\scshape cflux} & $F_{nthcomp}^{\dagger}$ & $3.55\pm 0.01$ &$4.21\pm 0.02$ \\
     \hline 
    & $\chi^{2}/dof$ &  $499/484$    &$350/315$  \\
    \hline
  \end{tabular}\label{table1} \\
  Flux values of {\tt nthcomp} component have been calculated in the $3-30$ keV energy band in the unit of $^{\dagger} 10^{-9}$ ergs cm$^{-2}$ s$^{-1}$. The parameter {\tt inp-type} of {\tt nthcomp} was fixed to $1$ (disk photon).
  \end{table*}

\subsection{Burst spectral analysis}
\begin{figure*}
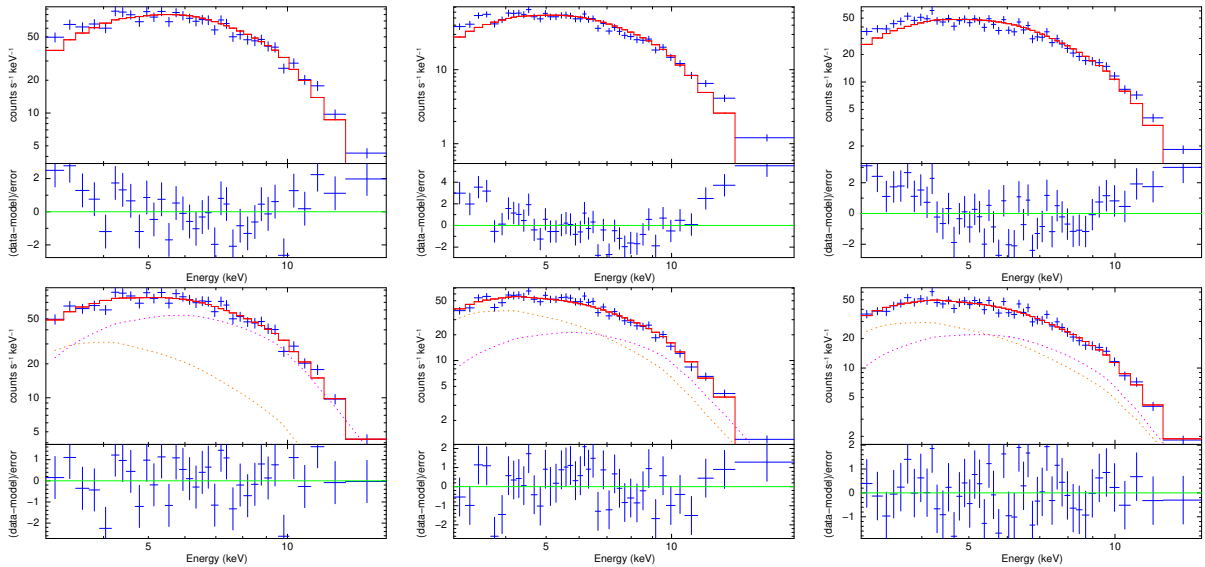

\includegraphics[scale=0.21, angle=-90]{fig6.eps}
\includegraphics[scale=0.21, angle=-90]{fig7.eps}
\includegraphics[scale=0.21, angle=-90]{fig8.eps}
\includegraphics[scale=0.21, angle=-90]{fig9.eps}
\includegraphics[scale=0.21, angle=-90]{fig10.eps}
\includegraphics[scale=0.21, angle=-90]{fig11.eps}
\caption{ {\bf Upper panel:} It shows the spectra around the peak of the burst for the segments S2 (20 s), S3(30 s), S4(40s), and the ratio of the data to the model in units of $\sigma$. The spectra are described by the model {\tt const*TBabs*bbodyrad}. The excesses at low and high energies are clearly visible.  {\bf Lower panel:} The spectra are fitted with the model {\tt const*TBabs*(bbodyrad+$f_{a}\times$ nthcomp)}. A variable persistent emission can fit the excesses observed in the low and high energies. The individual model components are shown: total model  (red), {\tt bbodyrad} (magenta), and {\tt $f_{a}\times$ nthcomp} (orange). } 
\label{Fig4}
\end{figure*}

We performed time-resolved spectroscopy of the long burst to investigate the evolution of the spectral parameters throughout the burst interval. We divided the entire burst interval into seven segments of different time intervals after the burst onset to cover the rising phase, peak, and long tail of the burst. The time intervals were chosen to be between $0 - 20$ s (segment 1, S1), $20 - 40$ s (S2), $40 - 70$ s (S3), $70 - 110$ s (S4), $110 - 160$ s (S5), $160 - 260$ s (S6), and $260 - 460$ s (S7) from the onset of the burst light curve (see Figure~\ref{Fig2}), following \citet{2020A&A...634A..58S}. After $460$ s, the light curve of the burst gradually flattens and eventually reaches the persistent level around $600$ s. We extracted the spectrum of each segment separately with a minimum of 100 counts in each spectral bin in the FPMA spectrum. To fit the burst spectra, we used an absorbed blackbody model {\tt TBabs*bbodyrad} with fixed absorption $0.2\times 10^{22}$ cm$^{-2}$. We found that a single blackbody radiation could not fit the burst spectrum. Although the burst spectra are dominated by a blackbody, excesses are evident at both low (soft X-ray excess) and high energy (hard X-ray deficit), see the upper panel of Figure~\ref{Fig4}. All the spectra extracted from different segments showed similar characteristics.\\

To take into account the residuals of the blackbody burst fit, we employed the enhanced persistent emission model. Initially, we performed the fit assuming that the persistent emission spectrum does not change during the burst, corresponding to the standard burst fitting technique (e.g. \citealt{2002A&A...382..947K, 2008ApJS..179..360G}). This model failed to fit the observed hard X-ray deficit and the soft X-ray excess for all the spectra. Later, we assumed that the amplitude of the persistent emission can change during the burst \citep{2013ApJ...772...94W} and adopted the following best-fit $f_{a}$ model composed of {\tt TBabs*(bbodyrad+$f_{a}\times$ nthcomp)}. The {\tt blackbody} and {\tt nthcomp} models account for the X-ray burst and persistent emissions during the burst, respectively. The parameters of the {\tt nthcomp} were fixed to the best-fit values of the preburst persistent emission taken from $400$ s time interval (reported in Table~\ref{table1}). The parameter $f_{a}$ is a free scaling factor that can be used to account for the variation of the persistent emission. If $f_{a} = 1$, this means that the amplitude of persistent emission during the burst is equal to the value before the burst. This model improved the fit and provided a reasonable description of all the burst spectra (see bottom panels of Figure~\ref{Fig4}). We obtained an acceptable fit for all the intervals of the burst with the model {\tt TBabs*(bbodyrad+$f_{a}\times$nthcomp)}, and it constrained all the free parameters simultaneously. We found that the best-fit values of $f_{a}$ is greater than $1$ for all segments, suggesting that the persistent emission typically increases during the burst. The best-fit parameter values for individual segments are quoted in Table~\ref{table2}. We used the {\tt XSPEC} model {\tt cflux} to determine the unabsorbed fluxes of different spectral components of the time-resolved spectra in the energy range $3 - 30$ keV during the burst (see Table~\ref{table2}). \\

\begin{table*}
   \centering
\caption{Fit results (burst emission): time-resolved spectroscopy of the burst observed in the source GS~1826-24. For the burst emission the model is:{\tt const*TBabs*(bbodyrad+$f_{a}\times$nthcomp)}} 
\begin{tabular}{|p{1.7cm}|p{1.8cm}|p{1.5cm}|p{1.5cm}|p{1.5cm}|p{1.5cm}|p{1.5cm}|p{1.5cm}|p{1.5cm}|}
    \hline
    Component     & Parameter (unit) & S1 (20s) & S2 (20s) & S3 (30s) & S4 (40s) & S5 (50s) & S6 (100s) & S7 (200s) \\
    \hline
    {\scshape tbabs}    & $N_{H}^{*}$ & $0.20 $(f)  & $0.20 $(f) &  $0.20 $(f) & $0.20 $(f)  & $0.20 $(f)  & $0.20 $(f)  & $0.20 $(f) \\
    {\scshape Mult. factor} & $f_{a}$ & $1.07_{-0.17}^{+0.15}$ & $ 1.54\pm 0.28$ & $1.88\pm 0.17$ & $1.44\pm 0.16$  &  $1.25\pm 0.15$ & $1.47\pm 0.10$  & $1.14\pm 0.04$ \\
    {\scshape cflux} &  $F_{f_{a}*nthcomp}^{\dagger}$ & $0.44\pm 0.01$ & $ 0.64\pm 0.01$ & $0.75\pm 0.02$ & $0.59\pm 0.01$  &  $0.52\pm 0.02$ & $0.59\pm 0.02$  & $0.49\pm 0.02$ \\
    {\scshape bbodyrad} & $kT_{bb}$ (keV) & $1.72_{-0.14}^{+0.26}$ & $2.10\pm 0.07$ & $2.05\pm 0.08$ & $1.71 \pm 0.05$  &  $1.51\pm 0.04$ & $1.39\pm 0.06$  & $1.09\pm 0.07$\\
    & Norm & $18_{-11}^{+16}$ & $89_{-14}^{+16} $ & $37_{-8}^{+10}$ & $69_{-14}^{+17}$  & $86_{-18}^{+19}$  &  $41_{-13}^{+14}$ & $71_{-14}^{+17}$\\
    & $R_{bb}$ (km) & $2.5_{-1.9}^{+2.2}$ & $5.5_{-2.1}^{+2.2} $ & $3.5_{-1.6}^{+1.8}$ & $4.8_{-2.1}^{+2.2}$  & $5.2_{-2.4}^{+2.5}$  &  $3.7_{-2.0}^{+2.1}$ & $4.8_{-2.1}^{+2.3}$\\
    {\scshape cflux} & $F_{bb}^{\dagger}$ & $0.14\pm 0.03$ & $ 1.70\pm 0.04$ & $0.66\pm 0.01$ & $0.55\pm 0.01$  &  $0.39\pm 0.01$ & $0.13\pm 0.01$  & $0.07\pm 0.01$ \\
    {\scshape cflux} &  $F_{tot}^{\dagger}$ & $0.58 \pm 0.01$ & $2.34 \pm 0.01$ & $1.41 \pm 0.01$ &$1.14 \pm 0.01$ & $0.91 \pm 0.01$ & $0.72 \pm 0.004$ & $0.55 \pm 0.003$ \\
     \hline 
    & $\chi^{2}/dof$ & $14/14$  & $35/32$ & $44/39$ & $41/45$  & $48/49$  & $63/72$  & $133/127$\\
    \hline
  \end{tabular}\label{table2} \\
  In unit of $^{*} 10^{22}\;\text{cm}^{-2}$. Flux values of different spectral component during the burst have been calculated in the $3-30$ keV energy band in the unit of $^{\dagger} 10^{-8}$ ergs cm$^{-2}$ s$^{-1}$. The parameters of {\tt nthcomp} were fixed to the persistent values (see Table~\ref{table1}).
\end{table*}

We tested other scenarios to fit the burst spectra apart from the $f_{a}$ model. First, we attempted to fit the burst spectra by the model {\tt TBabs*thcomp*bbodyrad}, considering that the burst photons could be affected by the corona/boundary layer \citep{2022ApJ...936...46C}. Here, {\tt thcomp} \citep{2020MNRAS.492.5234Z} is a more accurate version of {\tt nthcomp}, and the parameter values of this model are extracted from the persistent fit. However, this model failed to fit the residuals observed at low and high energies and provided a fit to the burst spectra that was significantly worse compared to the previous $f_{a}$ model. Second, we tried the model {\tt TBabs*(bbodyrad+powerlaw+nthcomp)} to fit the burst spectrum, in which the {\tt nthcomp} parameter values were fixed to the persistent emission \citep{2025ApJ...986...16J}. However, the current data failed to constrain all the free parameters simultaneously and could not produce conclusive results. Third, to fit the burst spectra, another combination of models like {\tt TBabs*(bbodyrad+\\relxillNS+nthcomp)} can be tested in which {\tt relxillNS} accounts for the reflection of the burst photons from the accretion disk \citep{2023A&A...670A..87L}. However, we could not implement this model because of the short exposure time of each segment of the burst spectrum and the lack of a priori knowledge of some important disk-related parameters. The above-mentioned facts restricted us from discussing these models. \\

\begin{figure}
\centering
\includegraphics[scale=0.52, angle=0]{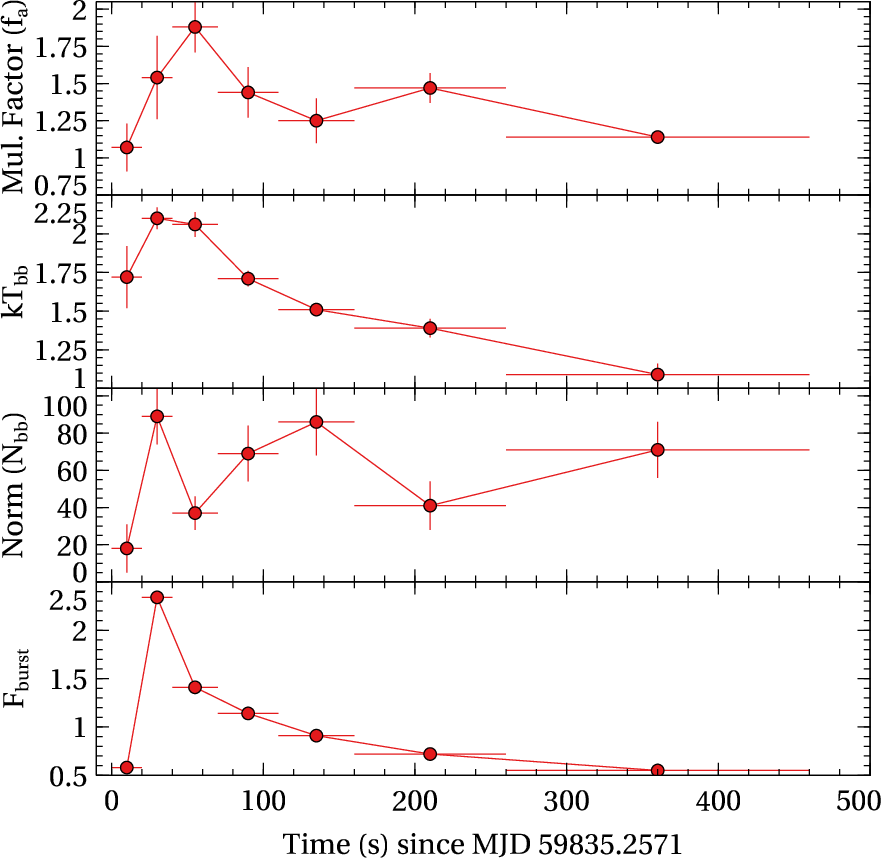}
\caption{The panels show variations of the best-fit spectral parameters: scaling factor of the pre-burst component ($f_{a}$), blackbody temperature ($kT_{bb}$), the blackbody normalization ($N_{bb}$), and flux ($F_{burst}$) during the burst for the best-fit model {\tt TBabs*(bbodyrad+$f_{a}\times$nthcomp)}. Horizontal bars indicate the width of the time bins, and the vertical bars are $1\sigma$ uncertainties. } 
\label{Fig5}
\end{figure}

\section{Discussion}
We performed a spectroscopic analysis of the long X-ray burst, observable for $\sim 600$ s, seen in the \nustar{} observation of the source GS~1826-24 to investigate the properties of the burst in detail. We first considered the spectral studies of the persistent emission of the source. We extracted persistent spectrum in the $3 - 30$ keV range from a time interval of 400 s just before the onset of the burst and a time interval of 2 ks before the burst, but from a different \nustar{} orbit. We modeled the persistent spectra using an absorbed cutoff power law ({\tt cutoffpl}) or a thermal Comptonization model ({\tt nthcomp}). Both models yielded a reasonably good description of the persistent emission in both cases. During this observation $3-30$ keV persistent flux was $F_{per}=(3.8\pm 0.2)\times 10^{-9}$ ergs cm$^{-2}$ s$^{-1}$, which correspond to a luminosity of $L_{per}\approx 1.62\times 10^{37}$ ergs s$^{-1}$ (assuming a source distance of $\sim 5.7$ kpc). This luminosity is roughly $9\%$ of the Eddington luminosity (where $L_{Edd}=1.8\times 10^{38}$ ergs s$^{-1}$ for a $1.4 M_{\odot}$ NS). The inferred persistent flux is consistent with the intermediate and soft state bolometric flux of the source \citep{2018MNRAS.474.1583J, 2020A&A...634A..58S, 2025ApJ...987..180G}. The local accretion rate per unit area can be calculated from the persistent luminosity following \citet{2008ApJS..179..360G}, which is $\dot{m}\approx 8.5\times 10^{3}$ g cm$^{-2}$ s$^{-1}$ for a NS of mass $1.4 M_{\odot}$ and radius 10 km in our case. Thus, the accretion rate onto the source was modest and exhibited a soft spectral state during this observation. \\

\begin{figure}
\centering
\includegraphics[scale=0.50, angle=0]{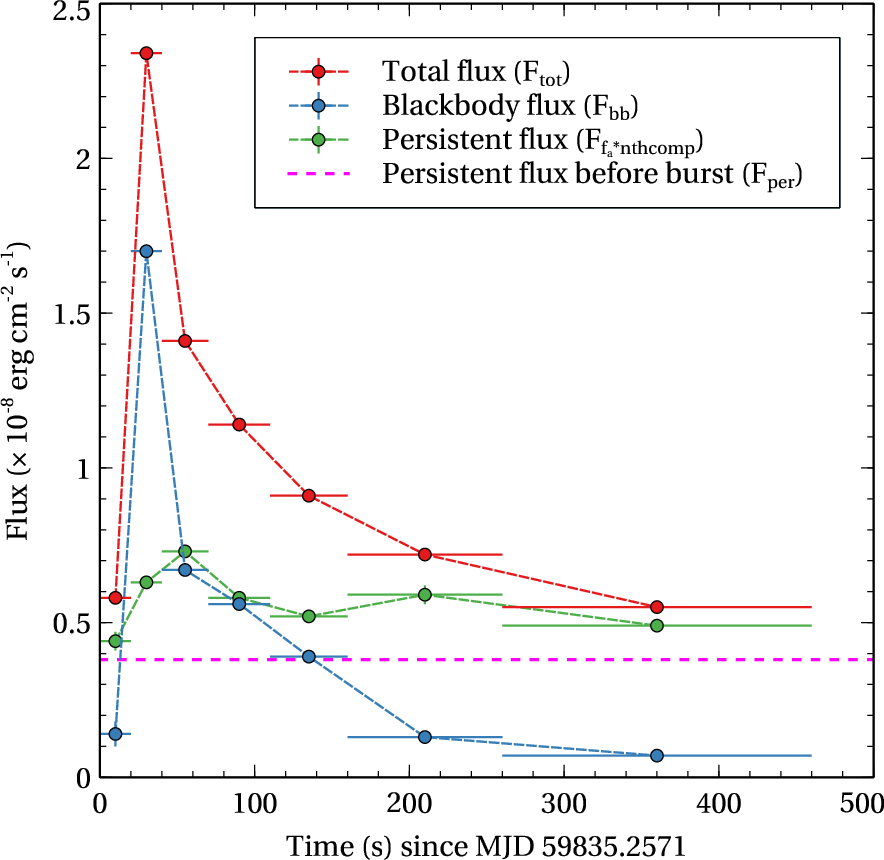}
\caption{The plot shows the variations of fluxes: total flux ($F_{tot}$), blackbody flux ($F_{bb}$), and the persistent flux $(F_{f_{a}*nthcomp})$ along the burst profile for the best-fit model {\tt TBabs*(bbodyrad+$f_{a}\times$nthcomp)}. The blackbody flux dominates at the peak of the burst, but after the peak, the persistent flux increases. The persistent flux ($F_{per}$) before the onset of the burst is shown by a dotted line.  } 
\label{Fig6}
\end{figure}

We extracted seven separate spectra along the burst profile, in the intervals 20 s, 20 s, 30 s, 40 s, 50 s, 100 s, and 200 s from the burst onset. We initially modeled the burst spectra using a simple absorbed blackbody model, but excesses are visible at low and high energies. To fit the time-resolved spectra of the burst properly, we used a model consisting of a blackbody plus a variable persistent emission component. We accounted for the excess flux in X-ray burst emission using the $f_{a}$ model of \citet{2013ApJ...772...94W}. Along the burst profile, the blackbody temperature decreases from the peak value $kT=2.10\pm 0.07$ keV to $kT=1.09\pm 0.06$ keV during the tail of the burst. The blackbody emitting radius ($R_{bb}$) is highest $5.5\pm 2.1$ km at the peak of the burst and becomes smaller as the flux decreases, but not in a regular fashion (see Figure~\ref{Fig5}). We did not observe any indication of photospheric radius expansion (PRE) at the peak of the burst from the time-resolved spectroscopy. The peak flux ($F_{peak}$) during the burst is $\approx 2.4\times 10^{-8}$ ergs cm$^{-2}$ s$^{-1}$. This flux corresponds to a luminosity of $\approx 9.7\times 10^{37}$ ergs s$^{-1}$, which is roughly $54\%$ of the $L_{Edd}$, indicating that the flux at the peak of the burst did not reach the Eddington limit. Additionally, we found that the best-fit values of $f_{a}$ are greater than $1$, suggesting that the persistent emission typically increases during the burst. The $f_{a}$ values rise at the burst start and quickly reach the maximum ($\sim 2$) within $\sim 50$ seconds, then they decrease slowly and return to around unity during the cooling tail (see top panel of Figure~\ref{Fig5}). This indicates that the persistent emission is rapidly enhanced due to the X-ray burst radiation and then returns to the preburst level during the cooling tail. To verify this, we further extracted a spectrum from a segment (after S7) of time interval $140$ s ($=600-460$ s) and fit in the same way as earlier for the spectra of segments S1...S7. The fit returned $f_{a}$ value of $1.02\pm 0.04$, thus confirming the above-mentioned fact. The enhancement in $f_{a}$ is perfectly consistent with the results obtained by \citet{2015ApJ...806...89J} after analyzing burst spectra seen by \rxte{}. We determined the unabsorbed $3 - 30$ keV blackbody flux ($F_{bb}$), the persistent flux ($F_{f_{a}*nthcomp}$), and showed their variation during the burst in the Figure~\ref{Fig6}. Around the burst peak, the blackbody flux supersedes the persistent flux; later on, the persistent flux starts dominating.\\

Our spectral analysis of the burst does not agree with the traditional assumption that the persistent, Comptonized emission does not vary during an X-ray burst. The time-resolved burst spectra from GS~1826-24 showed a significant departure from a pure thermal spectrum. During the entire burst interval, we observed an excess of soft and hard emission over the simple blackbody model. At the peak of the burst, the hard excess is less pronounced compared to the decaying phase (see top panel of Figure~\ref{Fig4}). We used a variable persistent emission to account for the excess flux in the X-ray burst emission. For this source, a drop in the hard X-ray emission, together with a soft X-ray excess with respect to the burst blackbody emission, has been observed earlier by \citet{2020A&A...634A..58S} from the \inte{} and \xmm{} observations. From the \rxte{} observations, hard X-ray shortages and enhancements of the persistent emission at soft X-rays during the burst have been observed from this source \citep{2015ApJ...806...89J}. However, The detection of excess burst emission has been observed for several other NS LMXBs: 4U~1820-30 \citep{2018ApJ...856L..37K}, Aql~X-1 \citep{2018ApJ...855L...4K}, SAX~J1808.4-3658 (\citealt{2013A&A...553A..83I, 2019ApJ...885L...1B}), 4U~1608-52 (\citealt{2019ApJ...883...61J, 2019JHEAp..24...23C}), 4U~1636-536 (\citealt{2022ApJ...935..154G, 2022MNRAS.509.3989K}), 4U~1728-34 \citep{2018ApJ...860...88B}, due to the increased sensitivity afforded by \nicer{}, {\it AstroSat}, and {\it Insight-HXMT}. This flux excess may indicate interactions between the burst emission and the accretion flow surrounding the NS. There may be various modes of interaction, but three processes are commonly considered to explain the excess flux. First, the burst emission may be reprocessed in the surface layers of the accretion disk and reflected into the line of sight \citep{2004ApJ...602L.105B}. In this case, the magnitude of the reflection component should approximately follow the burst intensity. Second, these excesses could be due to enhanced accretion rate onto the stellar surface through Poynting–Robertson (PR) drag \citep{1992ApJ...385..642W}, which corresponds to the change in persistent emission normalization \citep{2013ApJ...772...94W}. Third, The stellar surface could be covered by a Comptinizing medium, such that our view of the NS is partially obscured \citep{2014MNRAS.445.4218K}, thereby the flux excesses may be a result of the Comptonization \citep{2018ApJ...855L...4K} or scattering processes in the atmosphere \citep{1987ApJ...313..718R}. \\

It is important to understand how much of the observed X-ray emission is directly due to the thermonuclear process on the stellar surface and how much is added by the interaction with the stellar atmosphere. We, therefore, calculated the total flux along the X-ray burst profile and the contribution of the blackbody and the variable persistent emission separately. We found that the blackbody flux increases during the peak of the burst, and the contribution of the persistent emission is minimal (accounts for some $\sim 26\%$ of the total flux) during this interval. 
 It is expected because when an X-ary burst occurs in the NS envelope, the number of soft seed photons increases dramatically, enhancing the blackbody flux. However, the enhancement of the persistent flux is observed outside of the peak of the burst, and its contribution increases from $\sim 51\%$ to $\sim 89\%$ of the total flux during this decay phase (see Figure~\ref{Fig6}). We, therefore, observed a significant enhancement in the persistent emission during the burst. However, the underlying mechanism of such excess is not well understood. It has been observed that if the X-ray burst irradiates the accretion disk, it simultaneously results in both reflection and PR drag \citep{2020NatAs...4..541F}. We did not observe any reflection signature in the burst spectrum, which indicates that the first possibility of the excess emission during the burst may be ruled out. Thus, the enhancement in the persistent emission may be because of the PR effect, which temporarily increased the accretion rate onto the stellar surface \citep{1992ApJ...385..642W} or some other reprocessing of burst photons by the corona or the boundary layer. However, the data for the X-ray burst presented here are not sensitive enough to distinguish between these possibilities clearly. On the contrary, an apparent shortage of the persistent emission during the burst at hard X-rays (above 30 keV) indicates cooling of a corona \citep{2020A&A...634A..58S}. \\

The total duration of the burst, i.e., the time to evolve away from and return to the persistent state, was $\sim 600$ s in the \nustar{} FPMA/FPMB lightcurves. The burst has characteristics of both short-duration soft-state bursts and long-duration hard-state bursts \citep{2023ApJ...947...81Y, 2025ApJ...987..180G}. During this burst duration, the intensity of the source increased by at least $10\%$ (at the beginning and end point of the burst) compared to its persistent intensity level. We found that at about $\sim 15$ seconds after the burst onset, the flux reached $\sim 50\%$ of its peak flux. The burst exhibited a moderate rise time of $\sim 25$ s, which is the time spent between the start of the burst and the point at which $\sim 90\%$ of the peak burst intensity is reached. The bolometric flux ($0.1 - 100$ keV) during the burst is $\sim 1.1\times 10^{-8}$ ergs cm$^{-2}$ s$^{-1}$ which has been used to calculate the burst fluence ($f_{b}$). The value of $f_{b}$ was found to be $\approx 6.6\times 10^{-6}$ ergs cm$^{-2}$. The corresponding decay time of the burst is $\displaystyle\tau=f_{b}/F_{peak}\approx 282$ s. The measured fluence of the burst corresponds to a net burst energy release of $E_{burst}=4\pi d^{2}f_{b}=2.7\times 10^{40}$ erg (assuming a distance of $d=5.7$ kpc). The ignition depth ($y_{ing}$) is related to the energy release during the burst by the equation $E_{burst}=4\pi d^{2} y_{ing} Q_{nuc}/(1+z)$, where $Q_{nuc}$ is the energy release during nuclear reaction \citep{2003ApJ...595.1077C}. To calculate $y_{ing}$, we take an energy release of burning He to Ni of $Q_{nuc}\approx 10^{18}$ erg g$^{-1}$ (including losses due to neutrino emission), and assume a NS mass of $1.4 M_{\odot}$ and radius of $R=10$ km, which give a gravitational redshift of $1+z=1.31$. We found an ignitation depth $y_{ing}=2.8\times 10^{9}$ g cm$^{-2}$. At an accretion rate of $ 8.5\times 10^{3}$ g cm$^{-2}$ s$^{-1}$, the recurrence time corresponding to a column depth of $y_{ing}=2.8\times 10^{9}$ g cm$^{-2}$ is $\Delta t_{rec}=(y_{ign}/\dot{m})(1+z)\sim 5$ days. However, this $\Delta t_{rec}$ can not be compared to the earlier recurrence times as the current burst is observed when the source was not in an active clocked burster state. \\

To understand the burst fuel composition, one can estimate the $\alpha$ factor, which is the ratio of the persistent fluence between bursts to the total burst fluence of the burst, using the formula $\alpha=F_{per}\Delta t_{rec}/f_{b}$. In our case, there is no scope to calculate $\Delta t_{rec}$ from the observation, as we see only a single burst in this observation. Therefore, we verified the burst fuel composition in another way by determining the local accretion rate \citep{2008ApJS..179..360G}. A convenient unit of accretion rate is the Eddington rate. We, therefore, expressed the local accretion rate ($\dot{m}$) as a fraction of the local Eddington accretion rate ($\dot{m}_{Edd}$). We assumed the local accretion rate $\dot{m}_{Edd}=8.8\times 10^{4}\frac{1.7}{(X+1)}$ g cm$^{-2}$ s$^{-1}$, then we obtained $\dot{m}$ as a fraction of $\dot{m}_{Edd}$ for $X=0.7$ (for solar abundance) and $X=0$ (for pure helium). We found that $\dot{m}\sim 6\% \dot{m}_{Edd}$ and $\sim 10\% \dot{m}_{Edd}$ for $X=0$ and $X=0.7$, respectively. The value of $\dot{m}$ less than $10\%\dot{m}_{Edd}$ for $X=0$ supports the idea that these bursts were ignited in a helium-rich environment (see e.g., \citealt{2021ASSL..461..209G}) and such accretion rate is expected based on the burst behavior \citep{2008ApJS..179..360G}. Despite the uncertainty in the fuel composition, the properties of the bursts (i.e., longer burst duration) from GS~1826-24 strongly suggest burning of mixed H/He fuel by unstable He ignition \citep{1981ApJ...247..267F}. However, to study various burst properties in great detail, we need an uninterrupted observation of this source with broadband X-ray coverage.\\ 


\section{Data availability}
This research has made use of data obtained from the HEASARC, provided by NASA's Goddard Space Flight Center. The \nustar{} observational data set used in this work  (Obs. IDs $30801022002$), dated September 12, 2022, is in the public domain and is available on NASA's website https://heasarc.gsfc.nasa.gov. 
 
\section{Acknowledgements}
We want to thank the anonymous referee for his/her feedback and suggestions, which have significantly enhanced the quality of this paper. We would also like to thank Dr. Biplab Raychaudhuri for the productive discussions.
This research has made use of data and/or software provided by the High Energy Astrophysics Science Archive Research Centre (HEASARC). This research also has made use of the \nustar{} data analysis software ({\tt NuSTARDAS}) jointly developed by the ASI Space Science Data Center (SSDC, Italy) and the California Institute of Technology (Caltech, USA). ASM would like to thank Inter-University Centre for Astronomy and Astrophysics (IUCAA) for their facilities extended to him under their Visiting Associate Programme.

\def\apj{ApJ}
\def\apjl{ApJl}
\def\pasp{PASP} \def\mnras{MNRAS} \def\aap{A\&A} \def\physerp{PhR} \def\apjs{ApJS} \def\pasa{PASA}
\def\pasj{PASJ} \def\nat{Nature} \def\memsai{MmSAI} \def\araa{ARAA} \def\iaucirc{IAUC} \def\aj{AJ} \def\aaps{A\&AS} \def\ssr{SSR}
\def\iaucirc{iaucirc}
\bibliographystyle{unsrt}
\bibliography{aditya}

\begin{thebibliography}{58}
\providecommand{\natexlab}[1]{#1}
\providecommand{\url}[1]{\texttt{#1}}
\expandafter\ifx\csname urlstyle\endcsname\relax
  \providecommand{\doi}[1]{doi: #1}\else
  \providecommand{\doi}{doi: \begingroup \urlstyle{rm}\Url}\fi

\bibitem[{Arnaud}(1996)]{1996ASPC..101...17A}
K.~A. {Arnaud}.
\newblock {XSPEC: The First Ten Years}.
\newblock In G.~H. {Jacoby} and J.~{Barnes}, editors, \emph{Astronomical Data
  Analysis Software and Systems V}, volume 101 of \emph{Astronomical Society of
  the Pacific Conference Series}, page~17, 1996.

\bibitem[{Asai} et~al.(2015){Asai}, {Mihara}, {Matsuoka}, and
  {Sugizaki}]{2015PASJ...67...92A}
K.~{Asai}, T.~{Mihara}, M.~{Matsuoka}, and M.~{Sugizaki}.
\newblock {X-ray variability with spectral state transitions in NS-LMXBs
  observed with MAXI/GSC and Swift/BAT}.
\newblock \emph{\pasj}, 67\penalty0 (5):\penalty0 92, Oct. 2015.
\newblock \doi{10.1093/pasj/psv060}.

\bibitem[{Asai} et~al.(2022){Asai}, {Mihara}, and
  {Matsuoka}]{2022PASJ...74..974A}
K.~{Asai}, T.~{Mihara}, and M.~{Matsuoka}.
\newblock {Decades-long variations in NS-LMXBs observed with MAXI/GSC,
  RXTE/ASM, and Ginga/ASM}.
\newblock \emph{\pasj}, 74\penalty0 (4):\penalty0 974--990, Aug. 2022.
\newblock \doi{10.1093/pasj/psac049}.

\bibitem[{Ballantyne} and {Strohmayer}(2004)]{2004ApJ...602L.105B}
D.~R. {Ballantyne} and T.~E. {Strohmayer}.
\newblock {The Evolution of the Accretion Disk around 4U 1820-30 during a
  Superburst}.
\newblock \emph{\apjl}, 602:\penalty0 L105--L108, Feb. 2004.
\newblock \doi{10.1086/382703}.

\bibitem[{Bhattacharyya} et~al.(2018){Bhattacharyya}, {Yadav}, {Sridhar},
  {Verdhan Chauhan}, {Agrawal}, {Antia}, {Pahari}, {Misra}, {Katoch},
  {Manchanda}, and {Paul}]{2018ApJ...860...88B}
S.~{Bhattacharyya}, J.~S. {Yadav}, N.~{Sridhar}, J.~{Verdhan Chauhan}, P.~C.
  {Agrawal}, H.~M. {Antia}, M.~{Pahari}, R.~{Misra}, T.~{Katoch}, R.~K.
  {Manchanda}, and B.~{Paul}.
\newblock {Effects of Thermonuclear X-Ray Bursts on Non-burst Emissions in the
  Soft State of 4U 1728-34}.
\newblock \emph{\apj}, 860\penalty0 (2):\penalty0 88, June 2018.
\newblock \doi{10.3847/1538-4357/aac495}.

\bibitem[{Bildsten}(2000)]{2000AIPC..522..359B}
L.~{Bildsten}.
\newblock {Theory and observations of Type I X-Ray bursts from neutron stars}.
\newblock In S.~S. {Holt} and W.~W. {Zhang}, editors, \emph{Cosmic Explosions:
  Tenth AstroPhysics Conference}, volume 522 of \emph{American Institute of
  Physics Conference Series}, pages 359--369. AIP, June 2000.
\newblock \doi{10.1063/1.1291736}.

\bibitem[{Bult} et~al.(2019){Bult}, {Jaisawal}, {G{\"u}ver}, {Strohmayer},
  {Altamirano}, {Arzoumanian}, {Ballantyne}, {Chakrabarty}, {Chenevez},
  {Gendreau}, {Guillot}, and {Ludlam}]{2019ApJ...885L...1B}
P.~{Bult}, G.~K. {Jaisawal}, T.~{G{\"u}ver}, T.~E. {Strohmayer},
  D.~{Altamirano}, Z.~{Arzoumanian}, D.~R. {Ballantyne}, D.~{Chakrabarty},
  J.~{Chenevez}, K.~C. {Gendreau}, S.~{Guillot}, and R.~M. {Ludlam}.
\newblock {A NICER Thermonuclear Burst from the Millisecond X-Ray Pulsar SAX
  J1808.4-3658}.
\newblock \emph{\apjl}, 885\penalty0 (1):\penalty0 L1, Nov. 2019.
\newblock \doi{10.3847/2041-8213/ab4ae1}.

\bibitem[{Chen} et~al.(2013){Chen}, {Zhang}, {Zhang}, {Ji}, {Torres},
  {Kretschmar}, {Li}, and {Wang}]{2013ApJ...777L...9C}
Y.-P. {Chen}, S.~{Zhang}, S.-N. {Zhang}, L.~{Ji}, D.~F. {Torres},
  P.~{Kretschmar}, J.~{Li}, and J.-M. {Wang}.
\newblock {The Hard X-Ray Behavior of Aql X-1 during Type-I Bursts}.
\newblock \emph{\apjl}, 777\penalty0 (1):\penalty0 L9, Nov. 2013.
\newblock \doi{10.1088/2041-8205/777/1/L9}.

\bibitem[{Chen} et~al.(2019){Chen}, {Zhang}, {Zhang}, {Ji}, {Kong},
  {Santangelo}, {Qu}, {Lu}, {Li}, {Song}, {Xu}, {Cao}, {Chen}, {Liu}, {Bu},
  {Cai}, {Chang}, {Chen}, {Chen}, {Chen}, {Chen}, {Cui}, {Cui}, {Deng}, {Dong},
  {Du}, {Fu}, {Gao}, {Gao}, {Gao}, {Ge}, {Gu}, {Guan}, {Guo}, {Han}, {Huang},
  {Huo}, {Jia}, {Jiang}, {Jiang}, {Jin}, {Li}, {Li}, {Li}, {Li}, {Li}, {Li},
  {Li}, {Li}, {Li}, {Li}, {Liang}, {Liao}, {Liu}, {Liu}, {Liu}, {Liu}, {Lu},
  {Lu}, {Luo}, {Luo}, {Ma}, {Meng}, {Nang}, {Nie}, {Ou}, {Ren}, {Sai}, {Sun},
  {Tan}, {Tao}, {Tuo}, {Wang}, {Wang}, {Wang}, {Wang}, {Wang}, {Wen}, {Wu},
  {Wu}, {Wu}, {Xiao}, {Xiao}, {Xiong}, {Yang}, {Yang}, {Yang}, {Yang}, {Yi},
  {Yin}, {You}, {Zhang}, {Zhang}, {Zhang}, {Zhang}, {Zhang}, {Zhang}, {Zhang},
  {Zhang}, {Zhang}, {Zhang}, {Zhang}, {Zhang}, {Zhang}, {Zhang}, {Zhang},
  {Zhang}, {Zhao}, {Zhao}, {Zheng}, {Zhou}, {Zhou}, {Zhu}, and
  {Zhu}]{2019JHEAp..24...23C}
Y.~P. {Chen}, S.~{Zhang}, S.~N. {Zhang}, L.~{Ji}, L.~D. {Kong},
  A.~{Santangelo}, J.~L. {Qu}, F.~J. {Lu}, T.~P. {Li}, L.~M. {Song}, Y.~P.
  {Xu}, X.~L. {Cao}, Y.~{Chen}, C.~Z. {Liu}, Q.~C. {Bu}, C.~{Cai}, Z.~{Chang},
  G.~{Chen}, L.~{Chen}, T.~X. {Chen}, Y.~B. {Chen}, W.~{Cui}, W.~W. {Cui},
  J.~K. {Deng}, Y.~W. {Dong}, Y.~Y. {Du}, M.~X. {Fu}, G.~H. {Gao}, H.~{Gao},
  M.~{Gao}, M.~Y. {Ge}, Y.~D. {Gu}, J.~{Guan}, C.~C. {Guo}, D.~W. {Han},
  Y.~{Huang}, J.~{Huo}, S.~M. {Jia}, L.~H. {Jiang}, W.~C. {Jiang}, J.~{Jin},
  B.~{Li}, C.~K. {Li}, G.~{Li}, M.~S. {Li}, W.~{Li}, X.~{Li}, X.~B. {Li}, X.~F.
  {Li}, Y.~G. {Li}, Z.~W. {Li}, X.~H. {Liang}, J.~Y. {Liao}, G.~Q. {Liu}, H.~W.
  {Liu}, X.~J. {Liu}, Y.~N. {Liu}, B.~{Lu}, X.~F. {Lu}, Q.~{Luo}, T.~{Luo},
  X.~{Ma}, B.~{Meng}, Y.~{Nang}, J.~Y. {Nie}, G.~{Ou}, X.~Q. {Ren}, N.~{Sai},
  L.~{Sun}, Y.~{Tan}, L.~{Tao}, Y.~L. {Tuo}, C.~{Wang}, G.~F. {Wang},
  J.~{Wang}, W.~S. {Wang}, Y.~S. {Wang}, X.~Y. {Wen}, B.~Y. {Wu}, B.~B. {Wu},
  M.~{Wu}, G.~C. {Xiao}, S.~{Xiao}, S.~L. {Xiong}, J.~W. {Yang}, S.~{Yang},
  Y.~J. {Yang}, Y.~J. {Yang}, Q.~B. {Yi}, Q.~Q. {Yin}, Y.~{You}, A.~M. {Zhang},
  C.~L. {Zhang}, C.~M. {Zhang}, F.~{Zhang}, H.~M. {Zhang}, J.~{Zhang},
  T.~{Zhang}, W.~C. {Zhang}, W.~{Zhang}, W.~Z. {Zhang}, Y.~{Zhang}, Y.~F.
  {Zhang}, Y.~J. {Zhang}, Y.~{Zhang}, Z.~{Zhang}, Z.~L. {Zhang}, H.~S. {Zhao},
  X.~F. {Zhao}, S.~J. {Zheng}, D.~K. {Zhou}, J.~F. {Zhou}, Y.~{Zhu}, and Y.~X.
  {Zhu}.
\newblock {Insight-HXMT observation on 4U 1608-52: Evolving spectral properties
  of a bright type-I X-ray burst}.
\newblock \emph{Journal of High Energy Astrophysics}, 24:\penalty0 23--29, Nov.
  2019.
\newblock \doi{10.1016/j.jheap.2019.09.001}.

\bibitem[{Chen} et~al.(2022){Chen}, {Zhang}, {Ji}, {Zhang}, {Kong}, {Wang},
  {Chang}, {Peng}, {Qu}, and {Li}]{2022ApJ...936...46C}
Y.-P. {Chen}, S.~{Zhang}, L.~{Ji}, S.-N. {Zhang}, L.-D. {Kong}, P.-J. {Wang},
  Z.~{Chang}, J.-Q. {Peng}, J.-L. {Qu}, and J.~{Li}.
\newblock {Insight-HXMT Observation of 4U 1608-52: Evidence of Interplay
  between a Thermonuclear Burst and Accretion Environment}.
\newblock \emph{\apj}, 936\penalty0 (1):\penalty0 46, Sept. 2022.
\newblock \doi{10.3847/1538-4357/ac87a0}.

\bibitem[{Chenevez} et~al.(2016){Chenevez}, {Galloway}, {in 't Zand},
  {Tomsick}, {Barret}, {Chakrabarty}, {F{\"u}rst}, {Boggs}, {Christensen},
  {Craig}, {Hailey}, {Harrison}, {Romano}, {Stern}, and
  {Zhang}]{2016ApJ...818..135C}
J.~{Chenevez}, D.~K. {Galloway}, J.~J.~M. {in 't Zand}, J.~A. {Tomsick},
  D.~{Barret}, D.~{Chakrabarty}, F.~{F{\"u}rst}, S.~E. {Boggs}, F.~E.
  {Christensen}, W.~W. {Craig}, C.~J. {Hailey}, F.~A. {Harrison}, P.~{Romano},
  D.~{Stern}, and W.~W. {Zhang}.
\newblock {A Soft X-Ray Spectral Episode for the Clocked Burster, GS 1826-24 as
  Measured by Swift and NuStar}.
\newblock \emph{\apj}, 818\penalty0 (2):\penalty0 135, Feb. 2016.
\newblock \doi{10.3847/0004-637X/818/2/135}.

\bibitem[{Cocchi} et~al.(2011){Cocchi}, {Farinelli}, and
  {Paizis}]{2011A&A...529A.155C}
M.~{Cocchi}, R.~{Farinelli}, and A.~{Paizis}.
\newblock {BeppoSAX view of the NS-LMXB GS 1826-238}.
\newblock \emph{\aap}, 529:\penalty0 A155, May 2011.
\newblock \doi{10.1051/0004-6361/201016241}.

\bibitem[{Cumming}(2003)]{2003ApJ...595.1077C}
A.~{Cumming}.
\newblock {Models of Type I X-Ray Bursts from 4U 1820-30}.
\newblock \emph{\apj}, 595:\penalty0 1077--1085, Oct. 2003.
\newblock \doi{10.1086/377446}.

\bibitem[{Dickey} and {Lockman}(1990)]{1990ARA&A..28..215D}
J.~M. {Dickey} and F.~J. {Lockman}.
\newblock {H I in the Galaxy}.
\newblock \emph{\araa}, 28:\penalty0 215--261, 1990.
\newblock \doi{10.1146/annurev.aa.28.090190.001243}.

\bibitem[{Fragile} et~al.(2020){Fragile}, {Ballantyne}, and
  {Blankenship}]{2020NatAs...4..541F}
P.~C. {Fragile}, D.~R. {Ballantyne}, and A.~{Blankenship}.
\newblock {Interactions of type I X-ray bursts with thin accretion disks}.
\newblock \emph{Nature Astronomy}, 4:\penalty0 541--546, Jan. 2020.
\newblock \doi{10.1038/s41550-019-0987-5}.

\bibitem[{Fujimoto} et~al.(1981){Fujimoto}, {Hanawa}, and
  {Miyaji}]{1981ApJ...247..267F}
M.~Y. {Fujimoto}, T.~{Hanawa}, and S.~{Miyaji}.
\newblock {Shell flashes on accreting neutron stars and X-ray bursts}.
\newblock \emph{\apj}, 247:\penalty0 267--278, July 1981.
\newblock \doi{10.1086/159034}.

\bibitem[{Galloway} and {Keek}(2021)]{2021ASSL..461..209G}
D.~K. {Galloway} and L.~{Keek}.
\newblock {Thermonuclear X-ray Bursts}.
\newblock In T.~M. {Belloni}, M.~{M{\'e}ndez}, and C.~{Zhang}, editors,
  \emph{Timing Neutron Stars: Pulsations, Oscillations and Explosions}, volume
  461 of \emph{Astrophysics and Space Science Library}, pages 209--262, Jan.
  2021.
\newblock \doi{10.1007/978-3-662-62110-3_5}.

\bibitem[{Galloway} et~al.(2004){Galloway}, {Cumming}, {Kuulkers}, {Bildsten},
  {Chakrabarty}, and {Rothschild}]{2004ApJ...601..466G}
D.~K. {Galloway}, A.~{Cumming}, E.~{Kuulkers}, L.~{Bildsten}, D.~{Chakrabarty},
  and R.~E. {Rothschild}.
\newblock {Periodic Thermonuclear X-Ray Bursts from GS 1826-24 and the Fuel
  Composition as a Function of Accretion Rate}.
\newblock \emph{\apj}, 601\penalty0 (1):\penalty0 466--473, Jan. 2004.
\newblock \doi{10.1086/380445}.

\bibitem[{Galloway} et~al.(2008){Galloway}, {Muno}, {Hartman}, {Psaltis}, and
  {Chakrabarty}]{2008ApJS..179..360G}
D.~K. {Galloway}, M.~P. {Muno}, J.~M. {Hartman}, D.~{Psaltis}, and
  D.~{Chakrabarty}.
\newblock {Thermonuclear (Type I) X-Ray Bursts Observed by the Rossi X-Ray
  Timing Explorer}.
\newblock \emph{\apjs}, 179:\penalty0 360--422, Dec. 2008.
\newblock \doi{10.1086/592044}.

\bibitem[{Grefenstette} et~al.(2025){Grefenstette}, {Brumback}, {Buisson},
  {Ludlam}, {Mastroserio}, {Pike}, {Tomsick}, and {Yun}]{2025ApJ...987..180G}
B.~W. {Grefenstette}, M.~C. {Brumback}, D.~J.~K. {Buisson}, R.~M. {Ludlam},
  G.~{Mastroserio}, S.~N. {Pike}, J.~A. {Tomsick}, and S.~B. {Yun}.
\newblock {NuSTAR Observations of GS 1826-238 in the Extended Soft State}.
\newblock \emph{\apj}, 987\penalty0 (2):\penalty0 180, July 2025.
\newblock \doi{10.3847/1538-4357/addf3f}.

\bibitem[{G{\"u}ver} et~al.(2022){G{\"u}ver}, {Bostanc{\i}}, {Boztepe},
  {G{\"o}{\u{g}}{\"u}{\c{s}}}, {Bult}, {Kashyap}, {Chakraborty}, {Ballantyne},
  {Ludlam}, {Malacaria}, {Jaisawal}, {Strohmayer}, {Guillot}, and
  {Ng}]{2022ApJ...935..154G}
T.~{G{\"u}ver}, Z.~F. {Bostanc{\i}}, T.~{Boztepe},
  E.~{G{\"o}{\u{g}}{\"u}{\c{s}}}, P.~{Bult}, U.~{Kashyap}, M.~{Chakraborty},
  D.~R. {Ballantyne}, R.~M. {Ludlam}, C.~{Malacaria}, G.~K. {Jaisawal}, T.~E.
  {Strohmayer}, S.~{Guillot}, and M.~{Ng}.
\newblock {Burst-Disk Interaction in 4U 1636-536 as Observed by NICER}.
\newblock \emph{\apj}, 935\penalty0 (2):\penalty0 154, Aug. 2022.
\newblock \doi{10.3847/1538-4357/ac8106}.

\bibitem[{Harrison} et~al.(2013){Harrison}, {Craig}, {Christensen}, {Hailey},
  {Zhang}, {Boggs}, {Stern}, {Cook}, {Forster}, {Giommi}, {Grefenstette},
  {Kim}, {Kitaguchi}, {Koglin}, {Madsen}, {Mao}, {Miyasaka}, {Mori}, {Perri},
  {Pivovaroff}, {Puccetti}, {Rana}, {Westergaard}, {Willis}, {Zoglauer}, {An},
  {Bachetti}, {Barri{\`e}re}, {Bellm}, {Bhalerao}, {Brejnholt}, {Fuerst},
  {Liebe}, {Markwardt}, {Nynka}, {Vogel}, {Walton}, {Wik}, {Alexander},
  {Cominsky}, {Hornschemeier}, {Hornstrup}, {Kaspi}, {Madejski}, {Matt},
  {Molendi}, {Smith}, {Tomsick}, {Ajello}, {Ballantyne}, {Balokovi{\'c}},
  {Barret}, {Bauer}, {Blandford}, {Brandt}, {Brenneman}, {Chiang},
  {Chakrabarty}, {Chenevez}, {Comastri}, {Dufour}, {Elvis}, {Fabian}, {Farrah},
  {Fryer}, {Gotthelf}, {Grindlay}, {Helfand}, {Krivonos}, {Meier}, {Miller},
  {Natalucci}, {Ogle}, {Ofek}, {Ptak}, {Reynolds}, {Rigby}, {Tagliaferri},
  {Thorsett}, {Treister}, and {Urry}]{2013ApJ...770..103H}
F.~A. {Harrison}, W.~W. {Craig}, F.~E. {Christensen}, C.~J. {Hailey}, W.~W.
  {Zhang}, S.~E. {Boggs}, D.~{Stern}, W.~R. {Cook}, K.~{Forster}, P.~{Giommi},
  B.~W. {Grefenstette}, Y.~{Kim}, T.~{Kitaguchi}, J.~E. {Koglin}, K.~K.
  {Madsen}, P.~H. {Mao}, H.~{Miyasaka}, K.~{Mori}, M.~{Perri}, M.~J.
  {Pivovaroff}, S.~{Puccetti}, V.~R. {Rana}, N.~J. {Westergaard}, J.~{Willis},
  A.~{Zoglauer}, H.~{An}, M.~{Bachetti}, N.~M. {Barri{\`e}re}, E.~C. {Bellm},
  V.~{Bhalerao}, N.~F. {Brejnholt}, F.~{Fuerst}, C.~C. {Liebe}, C.~B.
  {Markwardt}, M.~{Nynka}, J.~K. {Vogel}, D.~J. {Walton}, D.~R. {Wik}, D.~M.
  {Alexander}, L.~R. {Cominsky}, A.~E. {Hornschemeier}, A.~{Hornstrup}, V.~M.
  {Kaspi}, G.~M. {Madejski}, G.~{Matt}, S.~{Molendi}, D.~M. {Smith}, J.~A.
  {Tomsick}, M.~{Ajello}, D.~R. {Ballantyne}, M.~{Balokovi{\'c}}, D.~{Barret},
  F.~E. {Bauer}, R.~D. {Blandford}, W.~N. {Brandt}, L.~W. {Brenneman},
  J.~{Chiang}, D.~{Chakrabarty}, J.~{Chenevez}, A.~{Comastri}, F.~{Dufour},
  M.~{Elvis}, A.~C. {Fabian}, D.~{Farrah}, C.~L. {Fryer}, E.~V. {Gotthelf},
  J.~E. {Grindlay}, D.~J. {Helfand}, R.~{Krivonos}, D.~L. {Meier}, J.~M.
  {Miller}, L.~{Natalucci}, P.~{Ogle}, E.~O. {Ofek}, A.~{Ptak}, S.~P.
  {Reynolds}, J.~R. {Rigby}, G.~{Tagliaferri}, S.~E. {Thorsett}, E.~{Treister},
  and C.~M. {Urry}.
\newblock {The Nuclear Spectroscopic Telescope Array (NuSTAR) High-energy X-Ray
  Mission}.
\newblock \emph{\apj}, 770:\penalty0 103, June 2013.
\newblock \doi{10.1088/0004-637X/770/2/103}.

\bibitem[{Hasinger} and {van der Klis}(1989)]{1989A&A...225...79H}
G.~{Hasinger} and M.~{van der Klis}.
\newblock {Two patterns of correlated X-ray timing and spectral behaviour in
  low-mass X-ray binaries}.
\newblock \emph{\aap}, 225:\penalty0 79--96, Nov. 1989.

\bibitem[{in't Zand} et~al.(2013){in't Zand}, {Galloway}, {Marshall},
  {Ballantyne}, {Jonker}, {Paerels}, {Palmer}, {Patruno}, and
  {Weinberg}]{2013A&A...553A..83I}
J.~J.~M. {in't Zand}, D.~K. {Galloway}, H.~L. {Marshall}, D.~R. {Ballantyne},
  P.~G. {Jonker}, F.~B.~S. {Paerels}, D.~M. {Palmer}, A.~{Patruno}, and N.~N.
  {Weinberg}.
\newblock {A bright thermonuclear X-ray burst simultaneously observed with
  Chandra and RXTE}.
\newblock \emph{\aap}, 553:\penalty0 A83, May 2013.
\newblock \doi{10.1051/0004-6361/201321056}.

\bibitem[{Iwakiri} et~al.(2018){Iwakiri}, {Serino}, {Mihara}, {Sugizaki},
  {Nakahira}, {Yatabe}, {Takao}, {Matsuoka}, {Ueno}, {Tomida}, {Ishikawa},
  {Sugawara}, {Isobe}, {Shimomukai}, {Kawai}, {Sugita}, {Yoshii}, {Tachibana},
  {Harita}, {Morita}, {Yoshida}, {Sakamoto}, {Kawakubo}, {Kitaoka},
  {Hashimoto}, {Tsunemi}, {Yoneyama}, {Negoro}, {Nakajima}, {Kawase},
  {Sakamaki}, {Maruyama}, {Ueda}, {Hori}, {Tanimoto}, {Oda}, {Morita},
  {Yamada}, {Tsuboi}, {Nakamura}, {Sasaki}, {Kawai}, {Sato}, {Yamauchi},
  {Hanyu}, {Hidaka}, {Kawamuro}, {Yamaoka}, and
  {Shidatsu}]{2018ATel11422....1I}
W.~{Iwakiri}, M.~{Serino}, T.~{Mihara}, M.~{Sugizaki}, S.~{Nakahira},
  F.~{Yatabe}, Y.~{Takao}, M.~{Matsuoka}, S.~{Ueno}, H.~{Tomida},
  M.~{Ishikawa}, Y.~{Sugawara}, N.~{Isobe}, R.~{Shimomukai}, N.~{Kawai},
  S.~{Sugita}, T.~{Yoshii}, Y.~{Tachibana}, S.~{Harita}, K.~{Morita},
  A.~{Yoshida}, T.~{Sakamoto}, Y.~{Kawakubo}, Y.~{Kitaoka}, T.~{Hashimoto},
  H.~{Tsunemi}, T.~{Yoneyama}, H.~{Negoro}, M.~{Nakajima}, T.~{Kawase},
  A.~{Sakamaki}, W.~{Maruyama}, Y.~{Ueda}, T.~{Hori}, A.~{Tanimoto}, S.~{Oda},
  T.~{Morita}, S.~{Yamada}, Y.~{Tsuboi}, Y.~{Nakamura}, R.~{Sasaki},
  H.~{Kawai}, T.~{Sato}, M.~{Yamauchi}, C.~{Hanyu}, K.~{Hidaka}, T.~{Kawamuro},
  K.~{Yamaoka}, and M.~{Shidatsu}.
\newblock {MAXI/GSC detections of a new superburst from GS 1826-238}.
\newblock \emph{The Astronomer's Telegram}, 11422:\penalty0 1, Mar. 2018.

\bibitem[{Iwata} et~al.(2025){Iwata}, {Ota}, {Kita}, {Yamasaki}, {Tsuchiya},
  {Nakano}, {Ichibakase}, {Takeda}, {Aoyama}, {Takahashi}, {Tamagawa}, {Enoto},
  {Iwakiri}, {Kitaguchi}, {Hu}, {Mihara}, and {NinjaSat
  Team}]{2025ATel17245....1I}
S.~{Iwata}, N.~{Ota}, T.~{Kita}, K.~{Yamasaki}, S.~{Tsuchiya}, Y.~{Nakano},
  M.~{Ichibakase}, T.~{Takeda}, A.~{Aoyama}, T.~{Takahashi}, T.~{Tamagawa},
  T.~{Enoto}, W.~{Iwakiri}, T.~{Kitaguchi}, C.~P. {Hu}, T.~{Mihara}, and
  {NinjaSat Team}.
\newblock {Return of the clocked burster GS 1826-238}.
\newblock \emph{The Astronomer's Telegram}, 17245:\penalty0 1, June 2025.

\bibitem[{Jaisawal} et~al.(2019){Jaisawal}, {Chenevez}, {Bult}, {in't Zand},
  {Galloway}, {Strohmayer}, {G{\"u}ver}, {Adkins}, {Altamirano}, {Arzoumanian},
  {Chakrabarty}, {Coopersmith}, {Gendreau}, {Guillot}, {Keek}, {Ludlam}, and
  {Malacaria}]{2019ApJ...883...61J}
G.~K. {Jaisawal}, J.~{Chenevez}, P.~{Bult}, J.~J.~M. {in't Zand}, D.~K.
  {Galloway}, T.~E. {Strohmayer}, T.~{G{\"u}ver}, P.~{Adkins}, D.~{Altamirano},
  Z.~{Arzoumanian}, D.~{Chakrabarty}, J.~{Coopersmith}, K.~C. {Gendreau},
  S.~{Guillot}, L.~{Keek}, R.~M. {Ludlam}, and C.~{Malacaria}.
\newblock {NICER Observes a Secondary Peak in the Decay of a Thermonuclear
  Burst from 4U 1608-52}.
\newblock \emph{\apj}, 883\penalty0 (1):\penalty0 61, Sept. 2019.
\newblock \doi{10.3847/1538-4357/ab3a37}.

\bibitem[{Jaisawal} et~al.(2025){Jaisawal}, {Chenevez}, {Strohmayer}, {Schatz},
  {in't Zand}, {G{\"u}ver}, {Altamirano}, {Arzoumanian}, and
  {Gendreau}]{2025ApJ...986...16J}
G.~K. {Jaisawal}, J.~{Chenevez}, T.~E. {Strohmayer}, H.~{Schatz}, J.~J.~M.
  {in't Zand}, T.~{G{\"u}ver}, D.~{Altamirano}, Z.~{Arzoumanian}, and K.~C.
  {Gendreau}.
\newblock {On the Origin of Spectral Features Observed during Thermonuclear
  X-Ray Bursts and in the Aftermath Emission of a Long Burst from 4U
  1820{\textendash}30}.
\newblock \emph{\apj}, 986\penalty0 (1):\penalty0 16, June 2025.
\newblock \doi{10.3847/1538-4357/adcc24}.

\bibitem[{Ji} et~al.(2013){Ji}, {Zhang}, {Chen}, {Zhang}, {Torres},
  {Kretschmar}, {Chernyakova}, {Li}, and {Wang}]{2013MNRAS.432.2773J}
L.~{Ji}, S.~{Zhang}, Y.~{Chen}, S.-N. {Zhang}, D.~F. {Torres}, P.~{Kretschmar},
  M.~{Chernyakova}, J.~{Li}, and J.-M. {Wang}.
\newblock {X-ray bursts as a probe of the corona: the case of XRB 4U 1636-536}.
\newblock \emph{\mnras}, 432\penalty0 (4):\penalty0 2773--2778, July 2013.
\newblock \doi{10.1093/mnras/stt625}.

\bibitem[{Ji} et~al.(2014){Ji}, {Zhang}, {Chen}, {Zhang}, {Torres},
  {Kretschmar}, and {Li}]{2014ApJ...782...40J}
L.~{Ji}, S.~{Zhang}, Y.~{Chen}, S.-N. {Zhang}, D.~F. {Torres}, P.~{Kretschmar},
  and J.~{Li}.
\newblock {The Hard X-Ray Shortages Prompted by the Clock Bursts in GS
  1826-238}.
\newblock \emph{\apj}, 782\penalty0 (1):\penalty0 40, Feb. 2014.
\newblock \doi{10.1088/0004-637X/782/1/40}.

\bibitem[{Ji} et~al.(2015){Ji}, {Zhang}, {Chen}, {Zhang}, {Torres},
  {Kretschmar}, {Kuulkers}, {Li}, and {Chang}]{2015ApJ...806...89J}
L.~{Ji}, S.~{Zhang}, Y.~{Chen}, S.-N. {Zhang}, D.~F. {Torres}, P.~{Kretschmar},
  E.~{Kuulkers}, J.~{Li}, and Z.~{Chang}.
\newblock {Diagnosing the Burst Influence on Accretion in the Clocked Burster
  GS 1826-238}.
\newblock \emph{\apj}, 806\penalty0 (1):\penalty0 89, June 2015.
\newblock \doi{10.1088/0004-637X/806/1/89}.

\bibitem[{Ji} et~al.(2018){Ji}, {Santangelo}, {Zhang}, {Ducci}, and
  {Suleimanov}]{2018MNRAS.474.1583J}
L.~{Ji}, A.~{Santangelo}, S.~{Zhang}, L.~{Ducci}, and V.~{Suleimanov}.
\newblock {Swift observations of GS 1826-238}.
\newblock \emph{\mnras}, 474\penalty0 (2):\penalty0 1583--1589, Feb. 2018.
\newblock \doi{10.1093/mnras/stx2908}.

\bibitem[{Kajava} et~al.(2014){Kajava}, {N{\"a}ttil{\"a}}, {Latvala},
  {Pursiainen}, {Poutanen}, {Suleimanov}, {Revnivtsev}, {Kuulkers}, and
  {Galloway}]{2014MNRAS.445.4218K}
J.~J.~E. {Kajava}, J.~{N{\"a}ttil{\"a}}, O.-M. {Latvala}, M.~{Pursiainen},
  J.~{Poutanen}, V.~F. {Suleimanov}, M.~G. {Revnivtsev}, E.~{Kuulkers}, and
  D.~K. {Galloway}.
\newblock {The influence of accretion geometry on the spectral evolution during
  thermonuclear (type I) X-ray bursts}.
\newblock \emph{\mnras}, 445\penalty0 (4):\penalty0 4218--4234, Dec. 2014.
\newblock \doi{10.1093/mnras/stu2073}.

\bibitem[{Kashyap} et~al.(2022){Kashyap}, {Ram}, {G{\"u}ver}, and
  {Chakraborty}]{2022MNRAS.509.3989K}
U.~{Kashyap}, B.~{Ram}, T.~{G{\"u}ver}, and M.~{Chakraborty}.
\newblock {Broad-band time-resolved spectroscopy of thermonuclear X-ray bursts
  from 4U 1636-536 using AstroSat}.
\newblock \emph{\mnras}, 509\penalty0 (3):\penalty0 3989--4007, Jan. 2022.
\newblock \doi{10.1093/mnras/stab2838}.

\bibitem[{Keek} et~al.(2018{\natexlab{a}}){Keek}, {Arzoumanian}, {Bult},
  {Cackett}, {Chakrabarty}, {Chenevez}, {Fabian}, {Gendreau}, {Guillot},
  {G{\"u}ver}, {Homan}, {Jaisawal}, {Lamb}, {Ludlam}, {Mahmoodifar},
  {Markwardt}, {Miller}, {Prigozhin}, {Soong}, {Strohmayer}, and
  {Wolff}]{2018ApJ...855L...4K}
L.~{Keek}, Z.~{Arzoumanian}, P.~{Bult}, E.~M. {Cackett}, D.~{Chakrabarty},
  J.~{Chenevez}, A.~C. {Fabian}, K.~C. {Gendreau}, S.~{Guillot},
  T.~{G{\"u}ver}, J.~{Homan}, G.~K. {Jaisawal}, F.~K. {Lamb}, R.~M. {Ludlam},
  S.~{Mahmoodifar}, C.~B. {Markwardt}, J.~M. {Miller}, G.~{Prigozhin},
  Y.~{Soong}, T.~E. {Strohmayer}, and M.~T. {Wolff}.
\newblock {NICER Observes the Effects of an X-Ray Burst on the Accretion
  Environment in Aql X-1}.
\newblock \emph{\apjl}, 855\penalty0 (1):\penalty0 L4, Mar. 2018{\natexlab{a}}.
\newblock \doi{10.3847/2041-8213/aab104}.

\bibitem[{Keek} et~al.(2018{\natexlab{b}}){Keek}, {Arzoumanian}, {Chakrabarty},
  {Chenevez}, {Gendreau}, {Guillot}, {G{\"u}ver}, {Homan}, {Jaisawal},
  {LaMarr}, {Lamb}, {Mahmoodifar}, {Markwardt}, {Okajima}, {Strohmayer}, and
  {in 't Zand}]{2018ApJ...856L..37K}
L.~{Keek}, Z.~{Arzoumanian}, D.~{Chakrabarty}, J.~{Chenevez}, K.~C. {Gendreau},
  S.~{Guillot}, T.~{G{\"u}ver}, J.~{Homan}, G.~K. {Jaisawal}, B.~{LaMarr},
  F.~K. {Lamb}, S.~{Mahmoodifar}, C.~B. {Markwardt}, T.~{Okajima}, T.~E.
  {Strohmayer}, and J.~J.~M. {in 't Zand}.
\newblock {NICER Detection of Strong Photospheric Expansion during a
  Thermonuclear X-Ray Burst from 4U 1820-30}.
\newblock \emph{\apjl}, 856\penalty0 (2):\penalty0 L37, Apr.
  2018{\natexlab{b}}.
\newblock \doi{10.3847/2041-8213/aab904}.

\bibitem[{Kuulkers} et~al.(2002){Kuulkers}, {Homan}, {van der Klis}, {Lewin},
  and {M{\'e}ndez}]{2002A&A...382..947K}
E.~{Kuulkers}, J.~{Homan}, M.~{van der Klis}, W.~H.~G. {Lewin}, and
  M.~{M{\'e}ndez}.
\newblock {X-ray bursts at extreme mass accretion rates from GX 17+2}.
\newblock \emph{\aap}, 382:\penalty0 947--973, Feb. 2002.
\newblock \doi{10.1051/0004-6361:20011656}.

\bibitem[{Kuulkers} et~al.(2003){Kuulkers}, {den Hartog}, {in't Zand},
  {Verbunt}, {Harris}, and {Cocchi}]{2003A&A...399..663K}
E.~{Kuulkers}, P.~R. {den Hartog}, J.~J.~M. {in't Zand}, F.~W.~M. {Verbunt},
  W.~E. {Harris}, and M.~{Cocchi}.
\newblock {Photospheric radius expansion X-ray bursts as standard candles}.
\newblock \emph{\aap}, 399:\penalty0 663--680, Feb. 2003.
\newblock \doi{10.1051/0004-6361:20021781}.

\bibitem[{Lewin} et~al.(1993){Lewin}, {van Paradijs}, and
  {Taam}]{1993SSRv...62..223L}
W.~H.~G. {Lewin}, J.~{van Paradijs}, and R.~E. {Taam}.
\newblock {X-Ray Bursts}.
\newblock \emph{\ssr}, 62:\penalty0 223--389, Sept. 1993.
\newblock \doi{10.1007/BF00196124}.

\bibitem[{Lu} et~al.(2023){Lu}, {Li}, {Pan}, {Yu}, {Chen}, {Ji}, {Ge}, {Zhang},
  {Qu}, {Song}, and {Falanga}]{2023A&A...670A..87L}
Y.~{Lu}, Z.~{Li}, Y.~{Pan}, W.~{Yu}, Y.~{Chen}, L.~{Ji}, M.~{Ge}, S.~{Zhang},
  J.~{Qu}, L.~{Song}, and M.~{Falanga}.
\newblock {Type I X-ray bursts' spectra and fuel composition from the atoll and
  transient source 4U 1730-22}.
\newblock \emph{\aap}, 670:\penalty0 A87, Feb. 2023.
\newblock \doi{10.1051/0004-6361/202244984}.

\bibitem[{Maccarone} and {Coppi}(2003)]{2003A&A...399.1151M}
T.~J. {Maccarone} and P.~S. {Coppi}.
\newblock {Spectral fits to the 1999 Aql X-1 outburst data}.
\newblock \emph{\aap}, 399:\penalty0 1151--1157, Mar. 2003.
\newblock \doi{10.1051/0004-6361:20021881}.

\bibitem[{Rodi} et~al.(2016){Rodi}, {Jourdain}, and
  {Roques}]{2016ApJ...817..101R}
J.~{Rodi}, E.~{Jourdain}, and J.~P. {Roques}.
\newblock {Hard X-Ray Tail Discovered in the Clocked Burster GS 1826-238}.
\newblock \emph{\apj}, 817\penalty0 (2):\penalty0 101, Feb. 2016.
\newblock \doi{10.3847/0004-637X/817/2/101}.

\bibitem[{Romani}(1987)]{1987ApJ...313..718R}
R.~W. {Romani}.
\newblock {Model Atmospheres for Cooling Neutron Stars}.
\newblock \emph{\apj}, 313:\penalty0 718, Feb. 1987.
\newblock \doi{10.1086/165010}.

\bibitem[{S{\'a}nchez-Fern{\'a}ndez} et~al.(2020){S{\'a}nchez-Fern{\'a}ndez},
  {Kajava}, {Poutanen}, {Kuulkers}, and {Suleimanov}]{2020A&A...634A..58S}
C.~{S{\'a}nchez-Fern{\'a}ndez}, J.~J.~E. {Kajava}, J.~{Poutanen},
  E.~{Kuulkers}, and V.~F. {Suleimanov}.
\newblock {Burst-induced coronal cooling in GS 1826-24. The clock wagging its
  tail}.
\newblock \emph{\aap}, 634:\penalty0 A58, Feb. 2020.
\newblock \doi{10.1051/0004-6361/201936599}.

\bibitem[{Strohmayer} and {Bildsten}(2006)]{2006csxs.book..113S}
T.~{Strohmayer} and L.~{Bildsten}.
\newblock \emph{{New views of thermonuclear bursts}}, pages 113--156.
\newblock Apr. 2006.

\bibitem[{Strohmayer} et~al.(2018){Strohmayer}, {Gendreau}, {Keek}, {Bult},
  {Mahmoodifar}, {Chakrabarty}, {Arzoumanian}, and {NICER Science
  Team}]{2018AAS...23133303S}
T.~E. {Strohmayer}, K.~C. {Gendreau}, L.~{Keek}, P.~{Bult}, S.~{Mahmoodifar},
  D.~{Chakrabarty}, Z.~{Arzoumanian}, and {NICER Science Team}.
\newblock {NICER Discovers mHz Oscillations and Marginally Stable Burning in GS
  1826-24}.
\newblock In \emph{American Astronomical Society Meeting Abstracts \#231},
  volume 231 of \emph{American Astronomical Society Meeting Abstracts}, page
  333.03, Jan. 2018.

\bibitem[{Swank} et~al.(1977){Swank}, {Becker}, {Boldt}, {Holt}, {Pravdo}, and
  {Serlemitsos}]{1977ApJ...212L..73S}
J.~H. {Swank}, R.~H. {Becker}, E.~A. {Boldt}, S.~S. {Holt}, S.~H. {Pravdo}, and
  P.~J. {Serlemitsos}.
\newblock {Spectral evolution of a long X-ray burst.}
\newblock \emph{\apjl}, 212:\penalty0 L73--L76, Mar. 1977.
\newblock \doi{10.1086/182378}.

\bibitem[{Tanaka}(1989)]{1989ESASP.296....3T}
Y.~{Tanaka}.
\newblock {Black-Holes in X-Ray Binaries - X-Ray Properties of the Galactic
  Black-Hole Candidates}.
\newblock In J.~{Hunt} and B.~{Battrick}, editors, \emph{Two Topics in X-Ray
  Astronomy, Volume 1: X Ray Binaries. Volume 2: AGN and the X Ray Background},
  volume~1 of \emph{ESA Special Publication}, page~3, Jan. 1989.

\bibitem[{Thompson} et~al.(2008){Thompson}, {Galloway}, {Rothschild}, and
  {Homer}]{2008ApJ...681..506T}
T.~W.~J. {Thompson}, D.~K. {Galloway}, R.~E. {Rothschild}, and L.~{Homer}.
\newblock {Deviations from the Flux-Recurrence Time Relationship in GS
  1826-238: Potential Transient Spectral Changes}.
\newblock \emph{\apj}, 681\penalty0 (1):\penalty0 506--514, July 2008.
\newblock \doi{10.1086/588723}.

\bibitem[{Ubertini} et~al.(1997){Ubertini}, {Bazzano}, {Cocchi}, {Natalucci},
  {Heise}, {Jager}, {in 't Zand}, {Muller}, {Smith}, {Celidonio}, {Coletta},
  {Ricci}, {Giommi}, {Ricci}, {Capalbi}, {Menna}, and
  {Rebecchi}]{1997IAUC.6611....1U}
P.~{Ubertini}, A.~{Bazzano}, M.~{Cocchi}, L.~{Natalucci}, J.~{Heise},
  R.~{Jager}, J.~{in 't Zand}, J.~M. {Muller}, M.~{Smith}, G.~{Celidonio},
  A.~{Coletta}, R.~{Ricci}, P.~{Giommi}, D.~{Ricci}, M.~{Capalbi}, M.~T.
  {Menna}, and S.~{Rebecchi}.
\newblock {GS 1826-238}.
\newblock \emph{\iaucirc}, 6611:\penalty0 1, Apr. 1997.

\bibitem[{Ubertini} et~al.(1999){Ubertini}, {Bazzano}, {Cocchi}, {Natalucci},
  {Heise}, {Muller}, and {in 't Zand}]{1999ApJ...514L..27U}
P.~{Ubertini}, A.~{Bazzano}, M.~{Cocchi}, L.~{Natalucci}, J.~{Heise}, J.~M.
  {Muller}, and J.~J.~M. {in 't Zand}.
\newblock {Bursts from GS 1826-238: A Clocked Thermonuclear Flashes Generator}.
\newblock \emph{\apjl}, 514\penalty0 (1):\penalty0 L27--L30, Mar. 1999.
\newblock \doi{10.1086/311933}.

\bibitem[{Verner} et~al.(1996){Verner}, {Ferland}, {Korista}, and
  {Yakovlev}]{1996ApJ...465..487V}
D.~A. {Verner}, G.~J. {Ferland}, K.~T. {Korista}, and D.~G. {Yakovlev}.
\newblock {Atomic Data for Astrophysics. II. New Analytic FITS for
  Photoionization Cross Sections of Atoms and Ions}.
\newblock \emph{\apj}, 465:\penalty0 487, July 1996.
\newblock \doi{10.1086/177435}.

\bibitem[{Walker}(1992)]{1992ApJ...385..642W}
M.~A. {Walker}.
\newblock {Radiation Dynamics in X-Ray Binaries. I. Type 1 Bursts}.
\newblock \emph{\apj}, 385:\penalty0 642, Feb. 1992.
\newblock \doi{10.1086/170969}.

\bibitem[{Wilms} et~al.(2000){Wilms}, {Allen}, and
  {McCray}]{2000ApJ...542..914W}
J.~{Wilms}, A.~{Allen}, and R.~{McCray}.
\newblock {On the Absorption of X-Rays in the Interstellar Medium}.
\newblock \emph{\apj}, 542:\penalty0 914--924, Oct. 2000.
\newblock \doi{10.1086/317016}.

\bibitem[{Worpel} et~al.(2013){Worpel}, {Galloway}, and
  {Price}]{2013ApJ...772...94W}
H.~{Worpel}, D.~K. {Galloway}, and D.~J. {Price}.
\newblock {Evidence for Accretion Rate Change during Type I X-Ray Bursts}.
\newblock \emph{\apj}, 772\penalty0 (2):\penalty0 94, Aug. 2013.
\newblock \doi{10.1088/0004-637X/772/2/94}.

\bibitem[{Yun} et~al.(2023){Yun}, {Grefenstette}, {Ludlam}, {Brumback},
  {Buisson}, {Mastroserio}, and {Pike}]{2023ApJ...947...81Y}
S.~B. {Yun}, B.~W. {Grefenstette}, R.~M. {Ludlam}, M.~C. {Brumback}, D.~J.~K.
  {Buisson}, G.~{Mastroserio}, and S.~N. {Pike}.
\newblock {Revealing the Spectral State Transition of the Clocked Burster, GS
  1826-238, with NuSTAR StrayCats}.
\newblock \emph{\apj}, 947\penalty0 (2):\penalty0 81, Apr. 2023.
\newblock \doi{10.3847/1538-4357/acb689}.

\bibitem[{Zdziarski} et~al.(1996){Zdziarski}, {Johnson}, and
  {Magdziarz}]{1996MNRAS.283..193Z}
A.~A. {Zdziarski}, W.~N. {Johnson}, and P.~{Magdziarz}.
\newblock {Broad-band {$\gamma$}-ray and X-ray spectra of NGC 4151 and their
  implications for physical processes and geometry.}
\newblock \emph{\mnras}, 283:\penalty0 193--206, Nov. 1996.

\bibitem[{Zdziarski} et~al.(2020){Zdziarski}, {Szanecki}, {Poutanen},
  {Gierli{\'n}ski}, and {Biernacki}]{2020MNRAS.492.5234Z}
A.~A. {Zdziarski}, M.~{Szanecki}, J.~{Poutanen}, M.~{Gierli{\'n}ski}, and
  P.~{Biernacki}.
\newblock {Spectral and temporal properties of Compton scattering by mildly
  relativistic thermal electrons}.
\newblock \emph{\mnras}, 492\penalty0 (4):\penalty0 5234--5246, Mar. 2020.
\newblock \doi{10.1093/mnras/staa159}.

\end{thebibliography}

\end{document}